\documentclass[sn-mathphys-num]{sn-jnl}

\usepackage{amsthm} 

\usepackage{csquotes}
\usepackage{epsfig}
\usepackage{epstopdf}
\usepackage{amssymb}
\usepackage{amscd}
\usepackage{amsmath}
\usepackage{amsfonts}
\usepackage[all,knot,poly]{xy}
\xyoption{arc}
\usepackage{latexsym}
\usepackage{graphicx}   
\usepackage{subcaption}
\usepackage[dvipsnames,usenames]{color}
\newtheorem{definition}{Definition}[section]

\newtheorem{proposition}{Proposition}[section]
\newtheorem{remark}{Remark}

\newcommand{\RVPP}{\textbf{RVPP}}
\newcommand{\IVPP}{\textbf{IVPP}}

\newcommand{\emphsay}[1]{\emph{\say{#1}}}
\newcommand{\EUconf}{\EU_\conf}

\newcommand{\Velrel}{{\Vect_\rel}}
\newcommand{\velrel}{{\vect_\rel}}

\newcommand{\derV}{{\der_{F}}}
\newcommand{\derdueV}{{\der^2_{F}}}

\newcommand{\mani}{\cM}
\newcommand{\Tmani}{{\TT\mani}}
\newcommand{\manin}{\cN}
\newcommand{\Tmanin}{{\TT\manin}}

\newcommand{\piola}{\BP}

\newcommand{\timeform}{\Btheta}
\newcommand{\FORME}{\Lambda}
\newcommand{\zerovec}{\mathbf{0}}

\newcommand{\PROJT}{\BR}
\newcommand{\PROJS}{\BP}

\newcommand{\image}{\mathbf{Im}}
\newcommand{\nucleo}{\mathbf{Ker}}

\newcommand{\area}{\cA}

\newcommand{\map}{\Bzeta}

\newcommand{\mapmoto}{(\map\push\moto)}
\newcommand{\refmap}{\Bp}

\newcommand{\Linspan}[1]{\mathbf{Span}\di{#1}}

\newcommand{\vect}{\Bv}
\newcommand{\vel}{\vect_{\moto}}
\newcommand{\Vect}{\BV}
\newcommand{\Vel}{\Vect_{\moto}}

\newcommand{\Velmap}{{\Vect_{\map\push\moto}}}
\newcommand{\velmap}{{\vect_{\map\push\moto}}}

\newcommand{\travel}{{\Bxi}}
\newcommand{\Veltravel}{{\Vect_\travel}}
\newcommand{\veltravel}{{\vect_\travel}}

\newcommand{\accel}{\Ba_{\moto}}

\newcommand{\extder}{\mathtt{d}}
\newcommand{\cof}{\mathbf{cof}}

\newcommand{\msec}{$\mathsection$}

\newcommand{\dvect}{\delta\Bv}
\newcommand{\proj}{\Bpi}

\newcommand{\normboxed}{\setlength\fboxsep{0.25cm}\boxed}

\newcommand{\immers}{\Bi}

\newcommand{\control}{\CC}
\newcommand{\pcontrol}{\partial\control}

\newcommand{\TjC}{{\Tj_\control}}

\newcommand{\equil}{\deform'}

\newcommand{\placement}{\Bp}

\newcommand{\flu}{\textsc{flu}}

\newcommand{\rel}{\textsc{rel}}
\newcommand{\ske}{\textsc{ske}}

\newcommand{\velflu}{{\vect_{\moto}^{\flu}}}
\newcommand{\velske}{{\vect_{\moto}^{\ske}}}

\newcommand{\ZEIT}{\cZ}

\newcommand{\shift}{\theta}

\newcommand{\EVE}{\cE}

\newcommand{\evento}{\Be}

\newcommand{\torsform}{\mathbb{\TT}}

\newcommand{\Lieder}{\cL}

\newcommand{\FEM}{\textbf{FEM}}
\newcommand{\MFI}{\textbf{MFI}}
\newcommand{\CM}{\textbf{CM}}
\newcommand{\ALE}{\textbf{ALE}}

\newcommand{\push}{{\uparrow}}
\newcommand{\pull}{{\downarrow}}
\newcommand{\forw}{{\Uparrow}}
\newcommand{\back}{{\Downarrow}}

\newcommand{\diverg}{\mathrm{div}}
\newcommand{\Diverg}{\mathrm{Div}}

\newcommand{\Bzero}{\mathbf{0}}

\newcommand{\refe}{\textsc{ref}}

\newcommand{\Lin}[1]{\mathbf{Lin\di{#1}}}

\newcommand{\VERT}{V}
\newcommand{\HORI}{H}
\newcommand{\TANG}{T}

\newcommand{\TEVE}{{\TANG\EVE}}
\newcommand{\VEVE}{{\VERT\EVE}}
\newcommand{\HEVE}{{\HORI\EVE}}

\newcommand{\TEVEs}{{(\TEVE)^*}}

\newcommand{\HEVEs}{{(\HEVE)^*}}
\newcommand{\Tj}{\cT}
\newcommand{\TTj}{{\TANG\Tj}}

\newcommand{\HTj}{{\HORI\Tj}}

\newcommand{\Tjs}{{\Tj_\map}}

\newcommand{\TjE}{{\Tj_\EVE}}

\newcommand{\HTjE}{{\HTj_\EVE}}

\newcommand{\TTjE}{{\TT\Tj_\EVE}}

\newcommand{\tE}{\ttt_\EVE}
\newcommand{\tEmap}{\ttt_\map}

\newcommand{\dtE}{\der\tE}

\newcommand{\conf}{{\BOmega}}
\newcommand{\pconf}{\partial\conf}
\newcommand{\Tconf}{{\TANG\conf}}

\newcommand{\confref}{\conf_{\refe}}
\newcommand{\pconfref}{\pconf_{\refe}}

\newcommand{\Fthrust}{\Bf_{\textsc{thr}}}

\newcommand{\timearrow}{\BZ}

\newcommand{\ext}{\textsc{ext}}
\newcommand{\inn}{\textsc{int}}
\newcommand{\dyn}{\textsc{dyn}}

\newcommand{\fext}{\Bf_\ext}
\newcommand{\finn}{\Bf_\inn}
\newcommand{\fdyn}{\Bf_\dyn}

\newcommand{\ID}[1]{\mathbf{id}_{#1}}

\newcommand{\TENS}{\textsc{Tens}}

\newcommand{\derF}{\der_F}

\newcommand{\stress}{\Bss}
\newcommand{\stressing}{\dot\stress}

\newcommand{\Celast}{\BPsi}

\newcommand{\Gelast}{\Xi}

\newcommand{\Gelasts}{\Gelast^*}

\newcommand{\es}{\Be\Bs}
\newcommand{\des}{\dot\Be\Bs}

\newcommand{\elstate}{\Be}

\newcommand{\el}{\textsc{el}}
\newcommand{\stretching}{\Beps}
\newcommand{\ela}{\stretching_{\el}}

\newcommand{\metric}{\Bg}

\newcommand{\FUN}{{\textsc{Fun}}}

\newcommand{\MIX}{{\textsc{Mix}}}

\newcommand{\COV}{{\textsc{Cov}}}
\newcommand{\CON}{{\textsc{Con}}}

\newcommand{\VOL}{\textsc{Vol}}

\newcommand{\dual}[1]{#1'}
\newcommand{\di}[1]{(#1)}

\newcommand{\coppia}[2]{(#1\,,#2)}
 
\newcommand{\Lbrack}[2]{[#1\,,#2]}

\newcommand{\CT}{\TT^*}

\newcommand{\massform}{\Bm}

\newcommand{\volform}{\Bmu}
\newcommand{\pvolform}{\partial\Bmu}
\newcommand{\volformg}{\volform_\metric}
\newcommand{\pvolformg}{\pvolform_\metric}

\newcommand{\EU}{S}

\newcommand{\unmezzotext}{\tfrac{1}{2}}

\newcommand{\bigdi}[1]{\big(#1\big)}
\newcommand{\Bigdi}[1]{\Big(#1\Big)}

\newcommand{\parder}[2]{\partial_{#1=#2}\,}

\newcommand{\nn}{n}

\newcommand{\sym}{\textrm{sym}}
\newcommand{\anti}{\textrm{skew}}

\newcommand{\cov}{\textsc{cov}}
\newcommand{\con}{\textsc{con}}

\newcommand{\BDelta}{\boldsymbol{\Delta}}

\newcommand{\Bomega}{\boldsymbol{\omega}}
\newcommand{\Boo}{\Bomega}
\newcommand{\BOmega}{\boldsymbol{\Omega}}

\newcommand{\BSigma}{\boldsymbol{\Sigma}}

\newcommand{\Bchi}{\boldsymbol{\chi}}
\newcommand{\BPi}{\boldsymbol{\Pi}}
\newcommand{\Bpi}{\boldsymbol{\pi}}

\newcommand{\Bss}{\boldsymbol{\sigma}}
\newcommand{\Bmu}{\boldsymbol{\mu}}

\newcommand{\Bvarphi}{\boldsymbol{\varphi}}

\newcommand{\Bzeta}{{\boldsymbol{\zeta}}}

\newcommand{\Btheta}{\boldsymbol{\theta}}

\newcommand{\Bxi}{\boldsymbol{\xi}}
\newcommand{\BXi}{\boldsymbol{\Xi}}

\newcommand{\BPsi}{\boldsymbol{\Psi}}
\newcommand{\Beps}{\boldsymbol{\epsilon}}

\newcommand{\Ba}{\mathbf{a}}
\newcommand{\Bb}{\mathbf{b}}
\newcommand{\Bc}{\mathbf{c}}

\newcommand{\Be}{\mathbf{e}}
\newcommand{\Bf}{\mathbf{f}}
\newcommand{\Bg}{\mathbf{g}}
\newcommand{\Bh}{\mathbf{h}}
\newcommand{\Bi}{\mathbf{i}}

\newcommand{\Bl}{\mathbf{l}}
\newcommand{\Bm}{\mathbf{m}}
\newcommand{\Bn}{\mathbf{n}}

\newcommand{\Bp}{\mathbf{p}}

\newcommand{\Br}{\mathbf{r}}
\newcommand{\Bs}{\mathbf{s}}
\newcommand{\Bt}{\mathbf{t}}
\newcommand{\Bu}{\mathbf{u}}
\newcommand{\Bv}{\mathbf{v}}
\newcommand{\Bw}{\mathbf{w}}
\newcommand{\Bx}{\mathbf{x}}

\newcommand{\BP}{\mathbf{P}}
\newcommand{\BR}{\mathbf{R}}

\newcommand{\BK}{\mathbf{K}}

\newcommand{\BE}{\mathbf{E}}
\newcommand{\BF}{\mathbf{F}}
\newcommand{\BM}{\mathbf{M}}

\newcommand{\BD}{\mathbf{D}}
\newcommand{\BI}{\mathbf{I}}
\newcommand{\BL}{\mathbf{L}}
\newcommand{\BX}{\mathbf{X}}

\newcommand{\BZ}{\mathbf{Z}}
\newcommand{\BW}{\mathbf{W}}
\newcommand{\BQ}{\mathbf{Q}}
\newcommand{\BT}{\mathbf{T}}

\newcommand{\BH}{\mathbf{H}}
\newcommand{\BV}{\mathbf{V}}

\newcommand{\BU}{\mathbf{U}}
\newcommand{\cP}{\mathcal{P}}
\newcommand{\cM}{\mathcal{M}}
\newcommand{\cA}{\mathcal{A}}
\newcommand{\cE}{\mathcal{E}}
\newcommand{\cB}{\mathcal{B}}

\newcommand{\cR}{\mathcal{R}}
\newcommand{\cL}{\mathcal{L}}
\newcommand{\cN}{\mathcal{N}}
\newcommand{\cV}{\mathcal{V}}
\newcommand{\cT}{\mathcal{T}}

\newcommand{\cF}{\mathcal{F}}

\newcommand{\cS}{\mathcal{S}}
\newcommand{\cX}{\mathcal{X}}

\newcommand{\cZ}{\mathcal{Z}}

\newcommand{\VV}{V}

\newcommand{\JJ}{J}
\newcommand{\TT}{T}

\newcommand{\CC}{C}
\newcommand{\HH}{H}

\newcommand{\pp}{p}

\newcommand{\ff}{f}

\newcommand{\kk}{k}

\newcommand{\scalar}[2]{{\langle}\kern.1em#1,#2\kern.1em{\rangle}}
\newcommand{\scalarC}[2]{{\langle}\kern.1em#1,#2\kern.1em{\rangle}_\conf}
\newcommand{\scalarT}[2]{{\langle}\kern.1em#1,#2\kern.1em{\rangle}_\TT}
\newcommand{\scalarTC}[2]{{\langle}\kern.1em#1,#2\kern.1em{\rangle}_{\TANG\conf}}

\newcommand{\inde}{\,\mathit{d}}
\newcommand{\integrale}[2]{\int_{#1}^{#2}}
\newcommand{\ointegrale}[2]{\oint_{#1}^{#2}}

\newcommand{\ttt}{t}
\newcommand{\sss}{s}
\newcommand{\mathscr}{\mathcal}

\newcommand{\equi}{\;\Longleftrightarrow\;}

\newcommand{\sub}[1]{{}_{\lower2pt\hbox{$\scriptstyle#1$}}}
\newcommand{\inv}[1]{#1^{-1}}

\newcommand{\punto}{\cdot}

\newcommand{\suchthat}{\mid}
\newcommand{\equaldef}{:=}
\newcommand{\perogni}{\quad\forall\,}

\newcommand{\set}[1]{\{#1\}}
\newcommand{\der}{d}

\newcommand{\tonde}[2]{(#1,#2)}

\newcommand{\persone}[1]{\textsc{#1}}
\newcommand{\Romano}{\persone{Romano}}

\newcommand{\Newton}{\persone{Newton}}

\newcommand{\Neumann}{\persone{Neumann}}
\newcommand{\Dirichlet}{\persone{Dirichlet}}
\newcommand{\Lions}{\persone{Lions}}
\newcommand{\Duvaut}{\persone{Duvaut}}

\newcommand{\Hilbert}{\persone{Hilbert}}
 
\newcommand{\Hill}{\persone{Hill}}

\newcommand{\Truesdell}{\persone{Truesdell}}
\newcommand{\Toupin}{\persone{Toupin}} 
\newcommand{\Noll}{\persone{Noll}} 

\newcommand{\Wang}{\persone{Wang}}
\newcommand{\Zaremba}{\persone{Zaremba}} 
\newcommand{\Jaumann}{\persone{Jaumann}}
\newcommand{\Oldroyd}{\persone{Oldroyd}}
\newcommand{\Bernstein}{\persone{Bernstein}}

\newcommand{\Green}{\persone{Green}}

\newcommand{\Riemann}{\persone{Riemann}}
\newcommand{\Leibniz}{\persone{Leibniz}}
\newcommand{\Lie}{\persone{Lie}}
\newcommand{\LeviCivita}{\persone{Levi-Civita}}

\newcommand{\Cauchy}{\persone{Cauchy}}
\newcommand{\Hooke}{\persone{Hooke}}
\newcommand{\Cosserat}{\persone{Cosserat}}
\newcommand{\Eringen}{\persone{Eringen}}
\newcommand{\Kirchhoff}{\persone{Kirchhoff}}
\newcommand{\Euler}{\persone{Euler}}
\newcommand{\Piola}{\persone{Piola}}

\newcommand{\Poisson}{\persone{Poisson}}

\newcommand{\Poincare}{\persone{Poincaré}}

\newcommand{\Banach}{\persone{Banach}}

\newcommand{\Whitney}{\persone{Whitney}}
\newcommand{\Legendre}{\persone{Legendre}}
\newcommand{\Euclid}{\persone{Euclid}}

\newcommand{\Navier}{\persone{Navier}}

\newcommand{\Stokes}{\persone{Stokes}}
\newcommand{\StVenant}{\persone{St.Venant}}
\newcommand{\Lagrange}{\persone{Lagrange}}
\newcommand{\Hamilton}{\persone{Hamilton}}

\newcommand{\Dalembert}{\persone{d'Alembert}}
\newcommand{\Bernoulli}{\persone{Bernoulli}}
\newcommand{\Varignon}{\persone{Varignon}}

\newcommand{\Simo}{\persone{Sim\'o}}

\newcommand{\Jacobi}{\persone{Jacobi}}

\newcommand{\Flugge}{\persone{Fl{\"u}gge}}
\newcommand{\Rivlin}{\persone{Rivlin}}
\newcommand{\Bertram}{\persone{Bertram}}
\newcommand{\Ryskin}{\persone{Ryskin}}
\newcommand{\Grassmann}{\persone{Grassmann}}
\newcommand{\Brezzi}{\persone{Brezzi}}
\newcommand{\Fortin}{\persone{Fortin}}
\newcommand{\Bilby}{\persone{Bilby}}
\newcommand{\Marsden}{\persone{Marsden}}
\newcommand{\Hughes}{\persone{Hughes}}

\newcommand{\Bruhns}{\persone{Bruhns}}
\newcommand{\Xiao}{\persone{Xiao}}
\newcommand{\Meyers}{\persone{Meyers}}
 
\newcommand{\Hencky}{\persone{Hencky}}
\newcommand{\Reuss}{\persone{Reuss}}
\newcommand{\Prandtl}{\persone{Prandtl}}
\newcommand{\Liu}{\persone{Liu}}
\newcommand{\Sampaio}{\persone{Sampaio}}
\newcommand{\Belytschko}{\persone{Belytschko}}
\newcommand{\Moran}{\persone{Moran}}

\newcommand{\Volterra}{\persone{Volterra}}
\newcommand{\Dantzig}{\persone{van Dantzig}}

\newcommand{\Coleman}{\persone{Coleman}}
\newcommand{\Mizel}{\persone{Mizel}}

\newcommand{\Epstein}{\persone{Epstein}}
\newcommand{\Gurtin}{\persone{Gurtin}}

\newcommand{\PodioGuidugli}{\persone{Podio-Guidugli}}
\newcommand{\Bigoni}{\persone{Bigoni}}

\newcommand{\Signorini}{\persone{Signorini}}

\newcommand{\Lee}{\persone{Lee}}

\newcommand{\Meshchersky}{\persone{Meshchersky}}

\newcommand{\Buquoy}{\persone{von Buquoy}}

\newcommand{\Nanson}{\persone{Nanson}}
\newcommand{\Fung}{\persone{Fung}}

\newcommand{\Duhem}{\persone{Duhem}}
\newcommand{\Malvern}{\persone{Malvern}}
\newcommand{\Ogden}{\persone{Ogden}}
\newcommand{\Merodio}{\persone{Merodio}}
\newcommand{\Nguyen}{\persone{Nguyen}}  
\newcommand{\Holzapfel}{\persone{Holzapfel}}
\newcommand{\Lacarbonara}{\persone{La Carbonara}}
\newcommand{\Mariano}{\persone{Mariano}}
\newcommand{\Galano}{\persone{Galano}}
\newcommand{\Lubarda}{\persone{Lubarda}}
\newcommand{\Asaro}{\persone{Asaro}}
\newcommand{\Oden}{\persone{Oden}}
\newcommand{\Reddy}{\persone{Reddy}}
\newcommand{\Fried}{\persone{Fried}}
\newcommand{\Anand}{\persone{Anand}}
\newcommand{\Salencon}{\persone{Salen\c{c}on}}
\newcommand{\Taroco}{\persone{Taroco}}
\newcommand{\Feijoo}{\persone{Feijóo}}
\newcommand{\Blanco}{\persone{Blanco}}

\newcommand{\Hesse}{\persone{Hesse}}

\newcommand{\Crisfield}{\persone{Crisfield}}
\newcommand{\Borst}{\persone{de Borst}}
\newcommand{\Ericksen}{\persone{Ericksen}}
\newcommand{\Bathe}{\persone{Bathe}}
\newcommand{\Man}{\persone{Man}}
\newcommand{\Fosdick}{\persone{Fosdick}}
\newcommand{\Temam}{\persone{Temam}}
\newcommand{\Miranville}{\persone{Miranville}}
\newcommand{\Freed}{\persone{Freed}}
\newcommand{\Rubin}{\persone{Rubin}}

\newcommand{\FL}[2]{\BF\Bl^{#1}_{#2}}

\newcommand{\moto}{\Bvarphi}
\newcommand{\lapse}{\alpha}
\newcommand{\blapse}{\beta}

\newcommand{\motoE}{\moto^\EVE}

\newcommand{\motoS}{\moto^\HH}
\newcommand{\motoZ}{\moto^\VV}

\newcommand{\vmoto}{\delta\Bvarphi}

\newcommand{\TEUconf}{\TT_\conf\EU}

\newcommand{\deform}{\BD}

\begin{document}

\title{Fundamental Topics in Continuum Mechanics: Grand Ideas, Errors \&\ Horrors}





\author[1]{\fnm{Giovanni} \sur{Romano}}\email{romano@unina.it} 

\author*[2]{\fnm{Raffaele} \sur{Barretta}}\email{rabarret@unina.it}

\affil[1]{Alma Mater: 
University of Naples Federico II, Naples, Italy}

\affil*[2]{Department of Structures for Engineering and Architecture, 
University of Naples Federico II, Via Claudio, 21 - 80125 Naples, Italy}


\date{Received: date / Accepted: date}


\abstract{
Shortly after the middle of the past century,
a comprehensive presentation of Continuum Mechanics 
was written under supervision of Clifford Ambrose \Truesdell\ III 
in two volumes of Siegfried \Flugge's  \emph{Handbuch der Physik},
a first in 1960 with Richard \Toupin\ 
on \emph{The Classical Field Theories} (\emph{the monster)},
including an Appendix on \emph{Tensor Analysis} 
by Jerald LaVerne \Ericksen,
and a second volume in 1965 with Walter \Noll\ 
on \emph{The Non-Linear Field Theories of Mechanics} (\emph{the monsterino)}.
Both nicknames are due to \Truesdell.
These contributions were gradually 
taken as turning points by the Mechanics Community worldwide,
due to completeness of analysis and profoundness of documentation.
Vastness of treatment acted however as a shield 
to careful reasoning on delicate  but basilar notions 
which, in the wake of some scholars of the XIX century, 
were taken to be worthy of belief and incorporated in the presentation
with a valuable historical background.
Lack of engineering perspective
didn't favour the necessary caution 
to be taken in facing a number of issues.
Scholars in Continuum Mechanics, fascinated by the monumental work 
conceived and carried out by \Truesdell\ and associates,
did not dare any accurate revision.
The analysis is here centred on unsatisfactory formulations
presently disseminated in literature by followers 
of 
\Truesdell's \emph{opus magnum}.
The geometric approach in 4D \Euclid\ spacetime
adopted here
is self-proposing even in the classical context,
providing clarity of  notions, methods and results
not achievable by the more familiar but less powerful 
and prone to confusing 3D treatment.
}

\keywords{
Motions in spacetime,
Material and spatial bundles,
Superposed Rigid Body Motions,
Change of observer,
Material frame indifference,
Continua with microstructure,
Alleged referential equilibria,
Rate elasticity,
Computational methods}



\maketitle

\section{Premise}
\label{sec: Premise}

Most theoretical presentations of Continuum Mechanics
(\CM)
are presently still developed in the \Euclid\ 3D spatial context,
with the time playing the role of evolution parameter.
Moreover, in the wake of treatments by
Clifford \Truesdell\ \cite{Truesdell1952,Truesdell1955},
\Truesdell\ \&\ Richard \Toupin\ \cite[\msec 210]{TruesdellToupin1960},
\Truesdell\ \&\ Walter \Noll\ \cite[\msec 43 A]{TruesdellNoll1965},
referential formulations of body and equilibrium are still proposed in spite of
alarm bells ringing after the diabolic deceits there involved were unveiled in
\cite{Treatise2023,Spacetime2023}.

Critical observations on the \emph{principle of equipresence},
formulated by Bernhard \Coleman\ and Victor \Mizel\ in \cite{ColemanMizel1964}
and adopted by \Truesdell\ \&\ \Noll\ in \cite{TruesdellNoll1965},
and on the \emph{principle of unification} 
stated by Ahmed  Cemal \Eringen\ in  \cite{Eringen1962},
were made by Ronald Samuel \Rivlin\ in \cite{Rivlin1972}
with the comment:

\emph{The relegation of physical considerations to a neglibible role 
in the formulation of physical theories in favor of arbitrary mathematical rules has, 
unfortunately, become too common a feature in modern nonlinear continuum theory. }

In reviewing a collection of \Noll's selected papers 
\cite{Noll1974},
dealing with the foundations of \CM,
critical remarks on effectiveness 
of adopted mathematical style and terminology
were also expressed by \Rivlin\ in \cite{Rivlin1976}.

Further considerations and comments made by \Rivlin\
in \cite{Rivlin1997} under the defiant title
\emph{Red Herrings and Sundry Unidentified Fish in Nonlinear Continuum Mechanics},
are relevant to the content and the intent of the present contribution
wherein 
other mechanical issues not highlighted by \Rivlin,
such as referential formulations of equilibrium and
the purported restriction imposed on constitutive relations
by \Noll's \emph{principle of isotropy of space} \cite{Noll1955},
renamed \emph{Material Frame Indifference} (\MFI)
by \Truesdell\ \cite[p. XII]{Noll2004}),
are detailedly discussed.

The critical remarks illustrated below target fundamental topics and
therefore push for a drastic revision of notions and results
in favour of physical significance and practical applicability.
Below is a long, but not exhaustive, list of books and articles, 
dating from the 1960s to the present, 
to testify the wide diffusion in the relevant literature 
of the 3D approach and of the alleged treatment 
of equilibrium in terms of a reference configuration.


\Fung\  \cite[\msec 16.3]{Fung1965},
\Truesdell\ \cite{Truesdell1966},
\Malvern\ \cite[Eq.(5.5.18)]{Malvern1969},
\Gurtin\ \cite{Gurtin1972},
\Wang\ \&\ \Truesdell\ \cite{WangTruesdell1973},
\Gurtin\ \cite[Ch.IX.27]{Gurtin1981},
\Oden\ \&\ \Reddy\ \cite[\msec.5.8]{OdenReddy1982},
\Marsden\ \&\ \Hughes\ \cite[Ch.5.4]{MarsdenHughes1983},
\Gurtin\ \cite[Ch.7]{Gurtin1983},
\Ogden\ \cite[\msec(3.4.2]{Ogden1984},
\Truesdell\ \cite{Truesdell1991},
\Crisfield\ \cite[Ch.(10.4)]{Crisfield1996},
\PodioGuidugli\ \cite[II.10]{Podio2000},
\Nguyen\ \cite[\msec{1.2.4}]{NguyenQS2000},
\Holzapfel\ \cite[Eq.(8.42)]{Holzapfel2000},
\Belytschko, \Liu\ W.-K., \Moran\ \cite[\msec 3.6]{Belytschko2001},
\Lubarda\ \cite[Ch.6]{Lubarda2002},
\Noll\ \cite{Noll2004},
\Asaro\ \&\ \Lubarda\ \cite[\msec.5.7]{AsaroLubarda2006},
\Man\ \&\ \Fosdick\ \cite{ManFosdick2005},
\Temam\ \&\ \Miranville\ \cite{TemamMiranville2005},
\Oden\ \cite[\msec 4.4]{Oden2006},
\Xiao, \Bruhns\ \&\  \Meyers\ \cite{XiaoBruhnsMeyers2006},
\Bertram\ \cite[Ch.3]{Bertram2008},
\Gurtin, \Fried\ \&\ \Anand\ \cite[Ch.24]{GurtinFriedAnand2010},
\Epstein\ \cite[Eq.(4.51)]{Epstein2010},
\Oden\ \cite[\msec{4.3}]{Oden2011},
\Borst\ \&\ al. \cite[\msec 3.4.1]{deBorst2012},
\Epstein\ \cite[Ch.6]{Epstein2012},
\Liu\ I.-S. \&\ \Sampaio\ R. \cite{LiuSampaio2012},
\Bigoni\ \cite[\msec{3.6}]{Bigoni2012},
\Lacarbonara\ \cite[Eq.(4.88)]{LacarbonaraWalter2013},
\Freed\ A.D. \cite{Freed2014},
\Mariano\ \&\ \Galano\ \cite{MarianoGalano2015}, 
\Salencon\ \cite{Salencon2016,Salencon2018},
\Taroco, \Blanco\ \&\ \Feijoo\ \cite[\msec 3.7]{Taroco2020},
\Merodio\ \&\ \Ogden\ \cite[Eq.(13)]{MerodioOgden2020},

However, when developing a geometric treatment, 
the manifold $\,\EVE\,$ of \Euclid\ spacetime events,
with $\,\dim\di\EVE=3+1\,$,
and its tangent bundle $\,\TEVE\,$
take readily the scene as natural mathematical setting,
even in classical mechanics.

Fundamental notions such as motion, velocity, acceleration,
change of observer
and definition of material and spatial fields,
can be properly introduced only in a spacetime context
wherein the geometric tools of \emph{time-projection}:
\begin{equation}
\tE:\EVE\mapsto\ZEIT
\label{fm: timeproj}
\end{equation}
onto the time-line $\,\ZEIT\,$, the 
\emph{observer time-arrows field}  
$\,\timearrow:\EVE\mapsto\TEVE\,$ and the
\emph{dynamical trajectory} $\,\TjE\subset\EVE\,$
\footnote{{\ }%
In 4D \Euclid\ spacetime, trajectories are bundles of non-intersecting lines
which describe the locus where body motions are detected.
On the contrary, in the 3D \Euclid\ space in which all spatial slices coalesce,
the trajectory is an intricate tangle of intersecting projected images.
Also, in 3D space there is no room for time-arrows,
see \msec\ref{sec: SMOS}.}
can be well-defined,
as described below in 
\msec\msec\ 
\ref{sec: Geopre},\ref{sec: SMOS},\ref{sec: motion}.

Inadequacy of mathematical modelling of the physical scenario
has contributed to realise a fertile ground 
for sneaky activities of deceptive devils.

Moreover, overbalance of formal mathematics versus 
engineering skill, made the flowering of trivial misstatements
possibly enter the scene, even hidden behind a somewhat pompous dressing.

Let us now come to the central contribution of this paper
aimed at presenting fundamental topics in Continuum Mechanics (\CM),
both on theoretical and computational sides,
under the comfortant umbrella of a physically sound geometric approach.

It is impressive that 
so many valuable scholars in \CM\ were imprinted
by authoritative writings of \Truesdell\ and associates,
to such an extent that danger signals definitely within reach
were not perceived.
These signals stem mainly from evident contrast to 
grand ideas on Mechanics
expressed by the original inventors of the Principles of this branch of Science.

Starting with Jacob \Bernoulli's foundational scientific discoveries and 
the brilliant ideas of his younger brother
Johann \Bernoulli, exposed in a 1715 letter to Pierre \Varignon\ 
\cite{Varignon1725},
we quote the extraordinary construction build up on contributions by
Daniel \Bernoulli\ \cite{BernoulliFamilie1969},
Jean-Baptiste Le Rond \Dalembert\ \cite{dAlembert1743},
Leonhard \Euler\ \cite{Euler1744},
Joseph-Louis \Lagrange\ \cite{Lagrange1788},
Siméon Denis \Poisson\ \cite{Poisson1811},
Augustin-Louis \Cauchy\ \cite{Cauchy1827,Cauchy1829},
George \Green\ \cite{Green1828,Green1839},
William Rowan \Hamilton\ \cite{Hamilton1835},
Carl Gustav Jacob \Jacobi\ \cite{Jacobi1841},
who in the epoch ranging from early XVIII to the middle XIX century
laid down the mathematical foundations of Continuum Mechanics and Dynamics.

Although some expert readers could feel difficulties
in following the formulation of definitions and properties of mechanical entities
in terms of basic differential geometric notions,
this powerful mathematical language is the one naturally apt to 
treat foundational aspects of Continuum Mechanics 
with clarity and precision otherwise not achievable.
\emph{Paris is well worth a mass!}
\footnote{{\ }%
Exclamation attributed to Henry IV the Great
on the occasion of his conversion
from Calvinism to Catholicism on July 25, 1593
before ascending to the throne of France.}

On the other hand, the presently usual approach is essentially algebraic in character,
with domain and range of fields and maps not explicitly specified.

This lack of description
is also responsible for unclear discussions and attempts of
revision, still lasting after more than half a century,
and even of misformulations which should finally be resolved, 
as in the auspices of this contribution.

\section{Geometric preliminaries}
\label{sec: Geopre}

In a nonlinear geometric framework,
the mathematical analysis 
is based on notions of
push by a flow and of parallel transport
over a differentiable manifold $\,\mani\,$.
Both are briefly exposed below,
while a full presentation can be found in 
\cite{Spivak1970,Abraham1983}.

A vector $\,\Bu_\Bx\in\TT_\Bx\mani\,$ tangent at $\,\Bx\in\mani\,$ 
is defined, for any smooth scalar field 
$\,\ff:\mani\mapsto\Re\,$, 
by the linear point-derivative $\,\der_\Bx\ff\,$
along a curve $\,\Bc:\Re\mapsto\mani\,$
parametrised so that $\,\Bc\di0=\Bx\,$:
\begin{equation}
\scalar{\der_\Bx\ff}{\Bu_\Bx}
\equaldef
\parder{\sss}{0}\di{\ff\circ\Bc}\di\sss\,.
\label{fm: tangvec}
\end{equation}

The tangent bundle $\,\Tmani\,$ to the manifold $\,\mani\,$
is the disjoint union of the family of tangent fibres
$\,\TT_\Bx\mani\,$, each labeled by the pertinent base point $\,\Bx\in\mani\,$.
The bundle projection%
\footnote{{\ }%
A \emph{projection} is a \emph{surjective submersion}
that is a surjective map such that its tangent map at each point is surjective too.
}
$\,\proj:\Tmani\mapsto\mani\,$ associates with each tangent
vector $\,\Bw_\Bx\in\TT_\Bx\mani\,$ the base point
$\,\Bx\in\mani\,$.
Tangent vector fields 
(in geometric terms sections of the tangent bundle)
are maps $\,\vect:\mani\mapsto\Tmani\,$
such that their composition 
with the bundle projection
$\,\proj\circ\vect:\mani\mapsto\mani\,$
is the identity in $\,\mani\,$.
This simply means that $\,\vect\di\Bx\in\TT_\Bx\mani\,$.

The tangent to a smooth map $\,\Bchi:\mani\mapsto\manin\,$  
between two manifolds $\,\mani\,$ and $\,\manin\,$
is the map $\,\TT\Bchi:\Tmani\mapsto\Tmanin\,$
which associates with any vector
$\,\BX\in\TT_\Bx\mani\,$, based at $\,\Bx\in\mani\,$
and tangent to a curve $\,\Bc:\Re\mapsto\mani\,$
at $\,\Bc\di0=\Bx\,$,
the corresponding vector $\,\TT\Bchi\punto\BX\,$
based at $\,\Bchi\di\Bx\in\manin\,$ 
and tangent to the curve 
$\,\Bchi\circ\Bc:\Re\mapsto\manin\,$.
A basic result due to 
Gottfried Wilhelm \Leibniz\ provides the rule for computing the tangent 
of the composition of two maps as chain of the single tangent maps
\cite[\msec 2.3]{Abraham1983}.

On a smooth manifold $\,\mani\,$
integration of a nowhere vanishing tangent vector field 
$\,\vect:\mani\mapsto\Tmani\,$ 
defines 
a regular flow $\,\FL\Bv\lambda:\mani\mapsto\mani\,$
such that
$\,\parder{\lambda}{0} \FL\Bv\lambda=\vect\,$.

The push of a scalar field 
$\,\alpha:\mani\mapsto\Re\,$ along a flow
is defined for all $\,\Bx\in\mani\,$ 
and $\,\lambda\in\Re\,$ by invariance:
\begin{equation}
 \di{\FL\Bv\lambda\push\alpha}_{\FL\Bv\lambda\di\Bx}\equaldef\alpha_\Bx
 \equi
 \FL\Bv\lambda\push\alpha=\alpha\circ\FL\Bv{-\lambda}
 \,.
\label{fm: scalarpush}
\end{equation}

The push of a vector field 
$\,\Bu:\mani\mapsto\Tmani\,$, along a smooth flow
$\,\FL\Bv\lambda:\mani\mapsto\mani\,$,
is defined at any point $\,\Bx\in\mani\,$
by means of the tangent functor $\,\TT\,$
\cite{Abraham1983}:h
\begin{equation}
 \di{\FL\Bv\lambda\push\Bu}_{\FL\Bv\lambda\di\Bx}
 \equaldef\di{\TT _\Bx\FL\Bv\lambda}\punto\Bu_\Bx
 \equi
 \FL\Bv\lambda\push\Bu=\di{\TT\FL\Bv\lambda}\circ\Bu\circ\FL{\Bv}{-\lambda}
 \,.
\label{fm: pushforward}
\end{equation}
The pull-back is the inverse correspondence 
$\,\FL\Bv\lambda\pull\Bu=\FL\Bv{-\lambda}\push\Bu\,$.

\medskip\goodbreak

In a manifold $\,\mani\,$ with a connection $\,\nabla\,$,
along any curve $\,\Bc:\Re\mapsto\mani\,$
to each parameter increment $\,\lambda\in\Re\,$
there corresponds a forward parallel transport 
$\,\di{\Bc_\lambda\forw\Bu_\Bx}_{\Bc\di\lambda}\,$ 
of any vector $\,\Bu_\Bx\in\TT_\Bx\mani\,$
and a backward transport:
\begin{equation}
\Bc_\lambda\back\equaldef\Bc_{-\lambda}\forw\,.
\label{fm: }
\end{equation}
A parallel transport independent of the curve joining start
and target points is said to be \emph{distant}.

A tensor field $\,\Bs:\mani\mapsto\TENS\di{\Tmani}\,$ 
is a multilinear function with vectors
(or dual covectors) as arguments, and living at points
(the value of the function at a point $\,\Bx\in\mani\,$  \cite{Spivak1970}
depends only on the values of the arguments at that point).

Push and parallel transport of tensor fields are defined by invariance
of their scalar values.

\Lie\ (or convective %
\footnote{\label{fn: Lie}{\ }%
Named \emph{Liesche ableitung},
after the Norwegian geometer Marius Sophus \Lie\ (1842--1899),
by the Dutch mathematician David \Dantzig\ \cite{Dantzig1954}.
}%
) derivatives $\,\Lieder_\Bv\,$
and parallel (covariant) derivatives $\,\nabla_\Bv\,$
of a tensor field $\,\Bs:\mani\mapsto\TENS\di{\Tmani}\,$%
\footnote{\label{fn: tens}{\ }%
In the tensor bundle $\,\TENS\di\Tmani\,$,
tensors at a point of the base manifold $\,\mani\,$
are multilinear real valued maps 
whose arguments are vectors or covectors 
at that point of $\,\mani\,$.
\emph{Covariant} tensors in $\,\COV\di\Tmani\,$ have vector arguments in $\,\Tmani\,$
while the arguments of \emph{contravariant} ones in $\,\CON\di\Tmani\,$ are covectors in the dual bundle $\,\CT\mani\,$.
Second order \emph{mixed} tensors in $\,\MIX\di\Tmani\,$ have vector-covector pairs
as arguments.
Scalar functions in $\,\FUN\di\Tmani\,$ are zeroth order tensors.
}
along a vector field $\,\Bv:\mani\mapsto\TT\mani\,$
are respectively defined by:
\begin{equation}
\setlength{\jot}{6pt}
\begin{aligned}
&\,\Lieder_\Bv\di\Bs
=\parder\lambda0\Bigdi{ \FL\Bv\lambda\pull\di{\Bs\circ\FL\Bv\lambda}}\,,
\\
&\,\nabla_\Bv\di\Bs
=\parder\lambda0\Bigdi{\FL\Bv\lambda\back\di{\Bs\circ\FL\Bv\lambda}}\,.
\end{aligned}
\label{fm: lieparalder}
\end{equation}

The parallel derivative $\,\nabla_\Bv\,$ is tensorial in 
the vector field $\,\Bv:\mani\mapsto\Tmani\,$ while the \Lie\ derivative 
$\,\Lieder_\Bv\,$ is not,
being dependent on the associated local flow.

For any smooth scalar field $\,\ff:\mani\mapsto\FUN\di\Tmani\,$
the \Lie\ \emph{bracket} of two tangent vector fields 
$\,\Bu,\Bv:\mani\mapsto\TT\mani\,$
is defined as the commutator
\cite{Spivak1970,HitchinNigel2003}:
\begin{equation}
\Lbrack{\Bv}{\Bu}\,\ff\equaldef(\Bv\Bu-\Bu\Bv)\,\ff\,.
\label{fm: Liebrack}
\end{equation}
Here the symbol $\,\Bu\ff:\mani\mapsto\FUN\di\Tmani\,$ denotes the derivative of the scalar field
$\,\ff:\mani\mapsto\FUN\di\Tmani\,$ along the vector field 
$\,\Bu:\mani\mapsto\TT\mani\,$.

A main result on \Lie\ differentiation states that:
\begin{equation}
\Lbrack{\Bv}{\Bu}=\Lieder_\Bv\di\Bu\,.
\label{fm: mainLie}
\end{equation}

Hence
$\,\Lieder_\Bu\di\Bu=\Lbrack{\Bu}{\Bu}=\Bzero\,$.

Exterior forms are alternating tensor fields
and exterior products between vectors fulfil the rules 
stated by Hermann Günther \Grassmann\ exterior algebra
\cite{Abraham1983}.

Let us now consider a chain
\footnote{\label{fn: chain}{\ }%
\emph{Chains} are formal sums of manifolds with signs depending
 on orientation compatibility \cite{Abraham1983}.
 A \emph{cochain} is dual to a chain according to 
 Vito \Volterra\ duality formula
 (a.k.a. Sir George Gabriel \Stokes\ formula), see Eq.\eqref{fm: Volterra} in \cite{Samelson2001}.}
 $\,\conf\,$ of compact manifolds in $\,\mani\,$
with boundary chain $\,\pconf\,$
and the dual co-chain of exterior forms.

\medskip\goodbreak

The notion of \emph{exterior derivative} of 
an exterior form
$\,\Boo\in\FORME^{(\nn-1)}\di\Tconf\,$,
with $\,\nn=\dim\di\conf\,$,
was introduced by Vito \Volterra\ 
with the following duality formula
where $\,\extder\Boo\in\FORME^\nn\di\Tconf\,$:
\begin{equation}
\integrale{\conf}{}\extder\Boo
=\ointegrale{\pconf}{}\Boo
\equi
\scalar{\extder\Boo}{\conf}=\scalar{\Boo}{\pconf}
\,.
\label{fm: Volterra}
\end{equation}

\goodbreak

From the definition $\,\partial\pconf=\Bzero\,$ 
(the boundary of a boundary is the null chain),
Eq.\eqref{fm: Volterra}, rewritten in terms of suitable duality pairing,
implies that the exterior differentiation is \emph{idempotent} too:
\begin{equation}
\extder\extder\Boo=\Bzero\quad \textrm{(null cochain)}\,.
\label{fm: zerochain}
\end{equation}

\section{Spacetime manifold and observers}
\label{sec: SMOS}

The proper context for the analysis of problems in Mechanics 
is the $4$D spacetime manifold of events $\,\EVE\,$
and its tangent bundle with \emph{projection} $\,\Bpi:\TEVE\mapsto\EVE\,$.

An \emph{observer} endows the tangent bundle $\,\TEVE\,$
with two geometric fields:
\begin{enumerate}
\itemsep=5pt
\item
A \emph{clock} one-form
$\,\timeform\in\FORME^1\di\TEVE:\EVE\mapsto\TEVEs\,$%
\footnote{\label{fn: dual}{\ }%
$\,\Lambda^\kk\,$ denotes the bundle of exterior forms of order $\,\kk\,$
and a superscript $\,*\,$ denotes duality.}
which is non-null and closed (i.e. with a vanishing exterior derivative):
\begin{equation}
\timeform\not=\zerovec\,,\quad
\extder\timeform=\zerovec\,.
\label{fm: timeformvan}
\end{equation}
\item
A nowhere vanishing
field of tangent \emph{time-arrows}
$\,\timearrow:\EVE\mapsto\TEVE\,$,
pointing towards the future and
named \emph{rigging} \cite{Friedman1983}
or \emph{observer field} \cite{Fecko2006,Fecko2013},
according to the suggestive language of physicists.
\end{enumerate}

A theorem by Vito \Volterra\ (a.k.a. Henri \Poincare\ Lemma)
\cite{Samelson2001}
ensures closedness and exactness
of an exterior form,
on a star shaped spacetime manifold $\,\EVE\,$,
are equivalent conditions \cite{objectivity2018,Treatise2023}.
Then, for the one-form $\,\timeform\in\FORME^1\di\TEVE\,$:
\begin{equation}
\extder\timeform=\Bzero\equi\timeform=\dtE\,.
\label{fm: timediff}
\end{equation}
The scalar potential:
\begin{equation}
\tE:\EVE\mapsto\ZEIT\,,
\label{fm: tp}
\end{equation}
is the \emph{time-projection}
onto the oriented $1$D \emph{time-axis} $\,\ZEIT\,$.%
\footnote{{\ }%
The axis $\,\ZEIT\,$ is identified with the real line $\,\Re\,$.
The letter $\,\ZEIT\,$ stands for \emph{Zeit} which is \emph{time} in German
\cite{Treatise2023}.}
It assigns a time instant $\,\tE\di\evento\in\ZEIT\,$ 
to each spacetime event $\,\evento\in\EVE\,$.

The horizontal tangent distribution
is composed of 
spacetime tangent vector fields $\,\Vect:\EVE\mapsto\TEVE\,$
fulfilling the condition 
\cite{Abraham1983}:
\begin{equation}
\Vect\in\nucleo\di\timeform\equi\scalar\timeform\Vect=0_\ZEIT\circ\tE\,.
\label{fm: }
\end{equation}

The spacetime manifold $\,\EVE\,$ is fibred into spatial slices $\,\EU\,$
which are integral manifolds of the kernel distribution $\,\nucleo\di\timeform\,$
of the clock one-form $\,\timeform\,$.

The \emph{clock} one-form
$\,\timeform\in\FORME^1\di\TEVE\,$
and the future pointing \emph{time-arrows field} 
$\,\timearrow:\EVE\mapsto\TEVE\,$
have a positive duality pairing
which, conveniently set to unity, is named the \emph{tuning}:%
\footnote{{\ }%
Covectors $\,\Ba^*\in\TEVEs:\TEVE\mapsto\Re\,$
are linear functionals.
The crochét $\,\scalar{\Ba^*}{\Ba}\,$ denotes the duality pairing
between covectors in $\,\Ba^*\in\TEVEs\,$
and tangent vectors $\,\Ba\in\TEVE\,$.
}%
\begin{equation}
\scalar{\timeform}{\timearrow}=1_\ZEIT\circ\tE\,.
\label{fm: tuning}
\end{equation}

A direct sum decomposition 
$\,\TEVE=\HEVE\oplus\VEVE\,$%
\footnote{{\ }%
$\,\VERT_\Be\EVE=\Lin{\timearrow_{\Be}}\,$ is pointwise the linear hull of $\,\timearrow\,$.}
holds
and tangent vectors $\,\Vect\in\TEVE\,$ are univocally split  
as sum $\,\Vect=\vect+\lambda\punto\timearrow\,$
with $\,\vect\in\HEVE\,$ and $\,\lambda\in\Re\,$.

The horizontåal bundle $\,\HEVE\,$ 
with $\,\dim\di\HEVE=3\,$
is endowed with a field
of metric tensors which are
symmetric and positive covariant tensors
$\,\metric\in\COV\di\HEVE\,$, so that each spatial slice 
$\,\EU\,$ is a \Riemann\ manifold.%
\footnote{{\ }%
Georg Friedrich Bernhard \Riemann\ (1826--1866), most prestigious 
German mathematican of the XIX century.
}

The dual bundle $\,(\HEVE)^*=\Lin\timearrow^\circ\,$%
\footnote{{\ }%
In the dual $\,\cX^*\,$ of a linear space $\,\cX\,$, the polar 
$\,\cL^\circ\subset\cX^*\,$ of a set 
$\,\cL\subset\cX\,$ is the linear subspace defined as
$\,\cL^\circ\equaldef\set{\Bu^*\in\cX^*\suchthat\scalar{\Bu^*}{\Bu}=0\perogni\Bu\in\cL}\,$.
}
is made of those covectors in $\,\TEVEs\,$ 
which vanish on time-arrows $\,\timearrow\in\TEVE\,$.
The bundle $\,(\HEVE)^*\,$ 
may be identified with the factor bundle
$\,\TEVEs/\di{\HEVE}^\circ\,$.

The notion of \emph{observer time-arrows} field
$\,\timearrow:\EVE\mapsto\TEVE\,$
puts in evidence a peculiarity of the
4D spacetime context when compared with the standard 3D spatial one
wherein the time plays just the role of ordering parameter,
with no room for time-arrows.%
\footnote{{\ }%
The decisive role of the \emph{observer time-arrows} field
$\,\timearrow:\EVE\mapsto\TEVE\,$ is evident when investigating
about effects of changes of observer
\cite{GCM2014,objectivity2018}.
}

In summary, the action of an observer consists of doubly foliating
the spacetime manifold $\,\EVE\,$ into:
\begin{enumerate}
\itemsep=5pt
\item[A.]
Leaves of \emph{isochronous} events ($3$D \emph{spatial slices}), 
i.e. integral manifolds of the kernel distribution $\,\nucleo\di\timeform\,$
of the \emph{clock} one-form $\,\timeform\,$.
\item[B.]
Lines of \emph{isotopic} events ($1$D \emph{spatial positions}).
\end{enumerate}

The $3$D \emph{spatial slices} and the $1$D spatial positions
 are mutually transversal, so that the
 \emph{tuning} Eq.\eqref{fm: tuning} is feasible.

Spacetime tensor fields of degree greater than zero
are \emph{horizontal} if they vanish
when any of their arguments is vertical, i.e. tangent to a time-line,
and are \emph{vertical}
if they vanish
when any of their arguments is thorizontal, i.e. tangent to a spatial slice.

\begin{definition}[Framing]\label{lm: framing}
The foliation performed by a spacetime
\emph{observer} is effectively described in geometrical terms 
by a \emph{framing}:%
\begin{equation}
\PROJT\equaldef\timeform\otimes\timearrow\,,
\label{fm: framing}
\end{equation}
a field of rank-one linear projectors on the \emph{time-arrows field}
$\,\timearrow:\EVE\mapsto\TEVE\,$,
according to the \emph{clock-rate} one-form
$\,\timeform\in\FORME^1\di\TEVE\,$.

Then, for all $\,\BX\in\TANG\EVE\,$:
\begin{equation}
\PROJT\punto\BX
=(\timeform\otimes\timearrow)\punto\BX=\scalar{\timeform}{\BX}\punto\timearrow\,.
\label{fm: projonzeta}
\end{equation}
\emph{Idempotency},
characteristic of linear projectors,
is equivalent to \emph{tuning}:
\begin{equation}
\PROJT\PROJT=\PROJT\equi\scalar\timeform\timearrow=1_\ZEIT\circ\tE\,.
\label{fm: tuningequiv}
\end{equation}
The \emph{horizontal} complementary projector defined by
$\,\PROJS=\BI-\PROJT\,$, is likewise \emph{idempotent}:
\begin{equation}
\setlength{\jot}{8pt}
\begin{aligned}
\PROJS\PROJS&\,=(\BI-\PROJT)(\BI-\PROJT)
=\BI-\PROJT-\PROJT+\PROJT\PROJT
=\BI-\PROJT=\PROJS\,.
\end{aligned}
\label{fm: name}
\end{equation}
Then:
\begin{equation}
\setlength{\jot}{6pt}
\left\{
\begin{aligned}
&\,\PROJS\PROJT=\PROJT\PROJS=\zerovec\,,
\\
&\,\image\di\PROJT=\nucleo\di\PROJS=\Linspan\timearrow\,,
\\
&\,\nucleo\di\PROJT=\,\image\di\PROJS=\nucleo\di\timeform\,.
\end{aligned}
\right.
\label{fm: projlist}
\end{equation}
\end{definition}

\section{Motion along the trajectory}
\label{sec: motion}

A primary example of powerfulness of the four-dimensional
spacetime representation is given by the mechanical notions of 
material trajectory $\,\Tj\,$ %
\footnote{{\ }%
The primitive physical character of the trajectory manifold
is not evidenced in standard treatments of \CM,
even in those with a geometric bias
\cite{Epstein2010,Epstein2012}
which prefer to embrace a deceptively simple potato-like picture
of a body and of purported reference placements.}
with immersion $\,\immers:\Tj\mapsto\EVE\,$  in spacetime
and of movements 
$\,\moto_\lapse:\TjE\mapsto\TjE\,$ 
along the immersed dynamical trajectory $\,\TjE=\immers\di\Tj\,$.

An \emph{observer} describes a
\emph{motion} $\,\moto\,$ of a material body %
\footnote{{\ }%
The identification of a material body and of its motion 
along the immersed trajectory is set up by means of a
specific interpretation of signals transmitted to an
observer even by non-mechanical phenomena
such as electro-magnetic fields, light, sound and heating waves.}
in the \Euclid\ spacetime $\,\EVE\,$ of events
as a one-parameter group of \emph{movements}:
\begin{equation}
\moto_\lapse:\TjE\mapsto\TjE\,.
\label{fm: movement}
\end{equation}
This group is commutative under the composition rule:
\begin{equation}
\moto_{(\lapse+\blapse)}=
\moto_\lapse\circ\moto_\blapse
=\moto_\blapse\circ\moto_\lapse\,,\perogni\lapse,\blapse\in\ZEIT\,.
\label{fm: comprule}
\end{equation}

Movements are automorphisms of the \emph{dynamical trajectory}  
$\,\TjE\subset\EVE\,$
required to fulfil simultaneity preservation
$\,\forall\;\lapse,\ttt\in\ZEIT\,$
according to the commutative diagram:
\begin{equation}
\kern-20pt
\begin{aligned}
\xymatrix@R=20pt@C=60pt{
{\TjE}
\ar[d]_{\tE}
\ar[r]^{\moto_\lapse}
&{\TjE}
\ar[d]^{\tE}
\\
{\ZEIT}
\ar[r]^{\shift_\lapse}
&{\ZEIT}
}
\end{aligned}
\;\equi\;
\left\{
\begin{aligned}
&\,\tE\circ\moto_\lapse=\shift_\lapse\circ\tE\,,
\\[3pt]
&\,\shift_\lapse\di\ttt\equaldef\ttt+\lapse\,.
\end{aligned}
\right.
\label{fm: diagmotouno}
\end{equation}

By applying the tangent functor $\,\TT\,$
to Eq.\eqref{fm: diagmotouno}, we infer invariance of the spatial bundle:
\begin{equation}
\der\tE\punto\TT\moto_\lapse=1_\ZEIT\punto\der\tE\,.
\label{fm: motionspainv}
\end{equation}

\goodbreak
This means vectors in a spatial slice are transformed 
by the tangent motion into vectors in another spatial slice.
The spacetime velocity of motion is defined by:
\begin{equation}
\Vel\equaldef\parder{\lapse}{0}\moto_\lapse:\TjE\mapsto\TT\TjE\,.
\label{fm: }
\end{equation}

Taking the derivative $\,\parder{\lapse}{0}\,$ in Eq.$\eqref{fm: diagmotouno}$,
we get:
\begin{equation}
\scalar{\dtE}{\Vel}=1_\ZEIT\circ\tE\,.
\label{fm: veltuning}
\end{equation}

Comparing Eq.\eqref{fm: veltuning}
with the \emph{tuning} property Eq.\eqref{fm: tuning},
we get a decomposition in space and time velocity components:
\begin{equation}
\Vel=\vel+\timearrow\,,
\label{fm: split}
\end{equation}
with the spatial component qualified by horizontality
$\,\scalar{\dtE}{\vel}=0\,$.

Moreover, defining of push:
\begin{equation}
\di{\moto\push\Vel}\circ\moto=\TT\moto\punto\Vel\,,
\label{fm: }
\end{equation}
tand aking the derivative $\,\parder{\lapse}{0}\,$ in Eq.\eqref{fm: comprule}, 
we infer spacetime velocity is pushed by the motion:
\begin{equation}
\moto\push\Vel=\Vel\,.
\label{fm: pushVel}
\end{equation}

Due to the spacetime split performed by an observer,
a spacetime motion
can be decomposed into a commutative chain of 
a horizontal-motion $\,\motoS\,$ (space) and a vertical-motion $\,\motoZ\,$ (time):%
\begin{equation}
\normboxed{\,
\vcenter{\halign{
\hfil$#$&$#$\hfil&$#$\hfil&$#$\hfil\cr
\motoE_\lapse&\,=\motoS_\lapse\circ\motoZ_\lapse
=\motoZ_\lapse\circ\motoS_\lapse
\,,\cr}}
\,}
\label{fm: motosplitframe}
\end{equation}

The two motions are envelopes of corresponding velocity fields:
\begin{equation}
\left\{
\setlength{\jot}{6pt}
\begin{aligned}
\vel&=\parder\lapse0\motoS_\lapse\,,
\\
\timearrow&=\parder\lapse0\motoZ_\lapse\,.
\end{aligned}
\right.
\label{fm: veldecomposition}
\end{equation}
Commutativity in Eq.\eqref{fm: motosplitframe}
is consistent with commutativity of addition
in Eq.\eqref{fm: split} and is equivalent to vanishing of the \Lie\ bracket
$\,\Lbrack{\vel}{\timearrow}\,$.

\section{Material and spatial fields}
\label{sec: msr}

Material and spatial fields are defined as follows 
\cite{GCM2014,objectivity2018,Spacetime2023}:
\begin{itemize}
\item
Material vector fields are sections of the material bundle,
that is maps from the base trajectory manifold $\,\Tj\,$
to the horizontal tangent bundle $\,\HTj\,$ to the trajectory $\,\Tj\,$.
The tangent bundle projection associates with each tangent vector 
the point where it is based on the trajectory.
Material tensor fields are multilinear maps on material vector and covector fields.
\item
Spatial vector fields are are sections of the spatial bundle
based on the trajectory manifold but taking values on the 
tangent bundle to the space-slice at the same time instant.
Spatial tensor fields are multilinear maps acting on spatial vector and covector fields.
\end{itemize}

In most present treatments of \CM\ 
following \cite{TruesdellNoll1965,Gurtin1981} 
referential vector fields based on a chosen reference placement  
and taking values on the reference tangent bundle are considered.

These referential fields are improperly labeled as material fields
and assumed to be in one-to-one correspondence with
spatial fields by means of a smooth placement map.

On the contrary,
according to the clear distinction set out in \cite{GCM2014}
and recalled above,
no one-to-one correspondence can be set up between material and spatial fields
according to the novel physically based definition.
Moreover:
\begin{itemize}
\item[a)]
Spatial vectors can be parallel transported along any line 
$\,\Bc:\Re\mapsto\EVE\,$
drawn in spacetime endowed with a connection.
The forward parallel transport is denoted by $\,\forw\,$ with inverse $\,\back\,$.
Invariance under parallel transport results in vani\-shing of the covariant derivative
$\,\nabla\,$ along the vector 
$\,\Bt=\parder{\lambda}{0}\Bc\di\lambda\,$
tangent to the line $\,\Bc:\Re\mapsto\EVE\,$,
defined by:
\begin{equation}
\setlength{\jot}{8pt}
\left\{
\begin{aligned}
\,\nabla_\Bt\di\vect
&\,\equaldef
\der_{\lambda\to0}\Bigdi{\back_{\lambda}\di{\vect\circ\Bc}\di\lambda}\,,
\\
&\,=\lim_{\lambda\to0}\frac{1}{\lambda}\Bigdi{\back_{\lambda}\di{\vect\circ\Bc}\di\lambda-\vect\di0}\,.
\end{aligned}
\right.
\label{fm: covder}
\end{equation}
Parallel transport to vectors based on \emph{alien} manifolds
outside the ambient spacetime is not feasible, 
since there is no available connection for resorting to.
\item[b)]
Material vectors are convected  
by the motion
$\,\moto_\lapse:\Tj\mapsto\Tj\,$
along the trajectory $\,\Tj\,$,
by push
$\,\push\,$ (with inverse pull $\,\pull\,$)
to get other material vectors,
as specified in Eq.\eqref{fm: pushforward}.
Invariance under push by the motion
results in vanishing of the \Lie\ 
(\emph{convective}) derivative defined, 
as in Eq.\eqref{fm: lieparalder}, by:
\begin{equation}
\setlength{\jot}{8pt}
\left\{
\begin{aligned}
\Lieder_{\Vel}\di{\overline\BV}\
&\,\equaldef
\parder{\lapse}{0}\di{\moto_\lapse\pull\overline\BV}
\\
&\,=\parder{\lapse}{0} \Bigdi{\TT\moto_{-\lapse}\punto\di{\overline\BV\circ\moto_{\lapse}}}\,,
\end{aligned}
\right.
\label{fm: conder}
\end{equation}
where $\,\Vel\equaldef\parder{\lapse}{0}\moto_\lapse:\Tj\mapsto\TTj\,$
is the spacetime motion velocity
and $\,\overline\BV:\Tj\mapsto\TTj\,$ any tangent vector field on the trajectory.
\end{itemize}

Parallel and convective derivatives of tensor fields are
defined by a formal application of \Leibniz\ rule,
resorting to invariance of scalar values 
both under parallel transport and under transformation by push
\cite{Treatise2023}.
The ensuing rules of convective derivation for covariant and contravariant tensor fields 
were given by James Gardner \Oldroyd\ in \cite{Oldroyd1950},
although in terms of components with inappropriate nomenclature.

\section{Superposed Rigid Body Motions}
\label{sec: SRBM}

The 3D treatment set out by
Clifford \Truesdell\ \&\ Walter \Noll\ 
in \cite{TruesdellNoll1965},
and adopted by
Morton Edward \Gurtin\  in
\cite[Ch.20, Eq.(1)]{Gurtin1981}
and with Eliot \Fried\ and Lallit \Anand\
in \cite[Ch.20.3, Eq.(20.10)]{GurtinFriedAnand2010},
is translated to spacetime context as follows.
\footnote{{\ }%
In \cite[\msec17, p.41]{TruesdellNoll1965} one can find the definition:
\emph{A change of frame is a one-to-one mapping
of space-time onto itself  such that distances, time intervals, and temporal order
are preserved.}.
All the subsequent analysis therein is confined to spatial isometric transformations.
}
\medskip
\begin{definition}[Euclid frame changes]\label{def: framechange}
Frame changes (a.k.a. \emph{changes of observer})
in \Euclid\ group
are diffeomorphisms $\,\map:\EVE\mapsto\EVE\,$
which leave invariant the fibers $\,\HORI_\Be\EVE\,$
of the horizontal bundle $\,\HORI\EVE\,$:%
\begin{equation}
\setlength{\jot}{8pt}
\left\{
\begin{aligned}
&\,\TT_\Be\map:\TT_\Be\EVE\mapsto\TT_\Be\EVE\,,
\\
&\,\TT_\Be\map:\HORI_\Be\EVE\mapsto\HORI_\Be\EVE\,,
\end{aligned}
\right.
\label{fm: spainv}
\end{equation}
and respect invariance of clock-rate and metric tensor:
\begin{equation}
\setlength{\jot}{8pt}
\left\{
\begin{aligned}
&\,\map\pull\extder\tE=\extder\tE\,, \quad\textrm{\Newton\ frame change}
\\
&\,\map\pull\metric=\metric\,.\qquad\quad\textrm{\Euclid\ frame change}\; (isometry)
\end{aligned}
\right.
\label{fm: NewtonEuclid}
\end{equation}
\end{definition}

Here $\,\metric:\HEVE\mapsto\HEVE^*\,$
(identified with $\,\metric:{\HEVE\otimes_\EVE\HEVE}\mapsto\Re\,$)
\footnote{{\ }%
The \Whitney\ bundle 
$\,\HEVE\otimes_\EVE\HEVE\,$
is the product bundle 
of two vector bundles over the same base manifold
$\,\EVE\,$
made of pairs of vectors in $\,\HEVE\,$ based at the same event.
}
is the symmetric and positive definite covariant metric tensor field
in the \Euclid\ spatial bundle.

The spatial restriction $\,\BQ:\HEVE\mapsto\HEVE\,$ 
of the tangent map $\,\TT\map:\TEVE\mapsto\TEVE\,$ 
preserves the base points,
as displayed by the next diagram
where $\,\Bpi:\HEVE\mapsto\EVE\,$
is the spatial bundle projection:
\begin{equation}
\begin{aligned}
\xymatrix@R=25pt@C=60pt{ 
{\HEVE}
\ar[d]^{\proj} 
\ar[r]^{\BQ}
&{\HEVE}
\ar[d]_{\proj} 
\\ 
{\EVE}
\ar[r]^{{\ID{\EVE}}} 
&{\EVE}
}
\end{aligned}
\quad\equi\quad
\proj\circ\,\BQ
=\proj\,.
\label{fm: basemoto}
\end{equation}

Defining the $\,\metric$-adjoint $\,\BQ^A\,$of $\,\BQ\,$
by the identity:
\begin{equation}
\metric\coppia{\BQ\Bu}{\BQ\Bv}=\metric\coppia{\BQ^A\BQ\Bu}{\Bv}
,\perogni\Bu,\Bv\in\HEVE\,,
\label{fm: giso}
\end{equation}
the properties of $\metric$-isometry,
and of spatial uniformity
according to \Euclid\ connection $\,\nabla\,$ by translation,
are expressed by:
\begin{equation}
\setlength{\jot}{8pt}
\left\{
\begin{aligned}
&\,\BQ^A=\inv\BQ\,, \quad\textrm{spatial isometry,}
\\
&\,\nabla\BQ=\Bzero\,, \qquad\textrm{spatial uniformity.}
\end{aligned}
\right.
\label{fm: iso}
\end{equation}
\medskip\goodbreak

In the treatment developed in \cite{TruesdellNoll1965},
a movement $\,\moto_\lapse:\Tj\mapsto\Tj\,$,
detected in the time lapse $\,\lapse\in\ZEIT\,$
by the an observer, after a frame change in the \Euclid\ group appears
to the transformed observer as a composed map
resulting from the superposition with the rigid transformation 
$\,\map:\EVE\mapsto\EVE\,$
fulfilling the conditions in 
Eqs.\eqref{fm: spainv}--\eqref{fm: iso}:
\begin{equation}
\moto^\map_\lapse\equaldef\map\circ\moto_\lapse:\Tj\mapsto\Tj_\map\equaldef\map\di\Tj\,.
\label{fm: superimposed}
\end{equation}
Accordingly, the tangent motion is said to transform according to the rule:
\begin{equation}
\TT\moto^\map=\BQ\punto\TT\moto\,.
\label{fm: tanmoveTN}
\end{equation}
Invariance under superposed rigid transformation
has also been recently resorted to and discussed
by Miles B. \Rubin\ in \cite[\msec 6.4]{Rubin2021}.

The deceit in this treatment
is most clearly revealed by a comparison with the
spacetime commutative diagram
of involved maps described in Eq.\eqref{fm: pushmoto} 
of the next \msec\ref{sec: FC}.

Let us remark once again that the spacetime approach is crucial 
to get a clear picture of the matter at hand.

\section{Frame change}
\label{sec: FC}

\medskip
\begin{definition}[Change of frame]\label{def: framechange}
In geometric terms, 
a \emph{change of frame} is defined
to be a diffeomorphism
$\,\map:\EVE\mapsto\EVE\,$
between two time-bundles 
$\,\tE:\EVE\mapsto\ZEIT\,$ and $\,\tEmap:\EVE\mapsto\ZEIT\,$
over the identity
$\,\ID{\ZEIT}:\ZEIT\mapsto\ZEIT\,$.
\end{definition}

A change of frame is effectively
expressed by the commutative diagram:
\begin{equation}
\begin{aligned}
\xymatrix@R=25pt@C=60pt{
\EVE
\ar@<.4ex>[r]^{\map}
\ar[d]^{\tE}
&\EVE
\ar@<.4ex>[l]^{\inv\map}
\ar[d]_{\tEmap}
\\
{\ZEIT}
\ar@{<->}[r]^{\ID{\ZEIT}}
\ar@{}[r]_{\phantom{\inv{\map}}}
&{\ZEIT}
}
\end{aligned}
\quad\equi\quad
\tEmap
=\tE\circ\inv\map
=\map\push\tE\,.
\label{fm: framechange}
\end{equation}

Taking the exterior differential of Eq.\eqref{fm: framechange},
by virtue of the commutativity:
\begin{equation}
\extder\di{\map\push\tE}=\map\push\extder\tE\,,
\label{fm: commut}
\end{equation}
we get the relation:
\begin{equation}
\extder\tEmap=\map\push\extder\tE\,.
\label{fm: }
\end{equation}

Classical mechanics allows only for \Newton\ frame changes 
which are characterised by covariance of clock rates, so that:
\begin{equation}
\extder\tEmap=\extder\tE\,.
\label{fm: newton}
\end{equation}
\medskip\goodbreak

A spacetime motion $\,\moto_\lapse:\TjE\mapsto\TjE\,$ is pushed 
by the change of frame $\,\map:\EVE\mapsto\EVE\,$
to a spacetime motion $\,\mapmoto_\lapse:\Tjs\mapsto\Tjs\,$,
with $\,\Tjs\equaldef\map\di\TjE\,$,
fulfilling the commutative diagram:%
\begin{equation}
\begin{aligned}
\xymatrix@R=25pt@C=60pt{ 
{\Tjs}
\ar[r]^{\mapmoto_\lapse}
&{\Tjs}
\\ 
{\TjE}
\ar[ur]^{\moto^\map_\lapse}
\ar[r]^{\moto_\lapse} 
\ar[u]^{\map} 
&{\TjE}
\ar[u]_{\map} 
}
\end{aligned}
\quad\equi\quad
\mapmoto_\lapse\circ\,\map
=\map\circ\,\moto_\lapse\,.
\label{fm: pushmoto}
\end{equation}

From Eq.\eqref{fm: pushmoto} the correct transformation rule for the tangent motion is inferred:
\begin{equation}
\TT\mapmoto\punto\TT\map=\TT\map\punto\TT\moto\,.
\label{fm: tanmot}
\end{equation}
In particular, for \Euclid\ frame changes:
\begin{equation}
\TT\mapmoto\punto\BQ=\BQ\punto\TT\moto\,.
\label{fm: EUtanmot}
\end{equation}

The pair of Eq.\eqref{fm: pushmoto} and \eqref{fm: EUtanmot}
provides the amendment to Eq.\eqref{fm: superimposed} and \eqref{fm: tanmoveTN}.

The many references in literature to \emph{superposed rigid body motions}
should be accordingly revised.

In fact, as expressed by Eq.\eqref{fm: superimposed}, the map 
$\,\map:\EVE\mapsto\EVE\,$,
which describes a spacetime change of frame as evaluated by the privileged observer,
being independent of the time lapse $\,\lapse\in\ZEIT\,$
is \emph{not} a movement
but an automorphic transformation in spacetime.

What is improperly called \emph{superposed rigid body motion}:
\begin{equation}
\moto^\map_\lapse=\map\circ\,\moto_\lapse:\TjE\mapsto\Tjs\,,
\label{fm: }
\end{equation}
is only an intermediate result, still in the middle of the ford.

Once the ford has been fully waded, the result appears 
to be the pushed motion depicted
in Eq.\eqref{fm: pushmoto}:
\begin{equation}
\mapmoto_\lapse=\map\circ\,\moto_\lapse\circ\,\inv\map:\Tjs\mapsto\Tjs\,.
\label{fm: }
\end{equation}

\begin{remark}\label{rem: MH}
The topic of superposed motion was dealt with
by \Marsden\ and \Hughes\ 
in a purely spatial context \cite{MarsdenHughes1983}. 
In Theorem 6.19 of \cite{MarsdenHughes1983}
they consider
the chain composition of a motion $\,\phi_{t}:\cB\mapsto\cS\,$ 
with velocity $\,\Bv_\phi:\cB\mapsto\TT\cS\,$,
of a body $\,\cB\,$ in the space $\,\cS\,$
and a superposed time dependent map $\,\xi_{t}:\cS\mapsto\cS\,$
with velocity
$\,\Bv_\xi:\cB\mapsto\TT\cS\,$.
Applied to the composition $\,\xi_{t}\circ\phi_{t}:\cB\mapsto\cS\,$,
\Leibniz\ rule gives the velocity 
$\,\Bv_{\xi\circ\phi}:\cB\mapsto\TT\cS=\Bv_{\xi}+\TT\xi\punto\Bv_{\phi}\,$
but  they write 
$\,\xi\push\Bv_{\phi}=\TT\xi\punto\Bv_{\phi}\circ\inv{\xi}\,$ in place of $\,\TT\xi\punto\Bv_{\phi}\,$.
This flaw, together with the unfeasible derivation  
$\,\parder{s}{t}\xi\push\BT\,$
and an obscure simplification in the third-last row of their proof,
are decisive in leading to the mistaken hasty conclusion:
"objective tensors have objective Lie-derivative."
However in \cite{MarsdenHughes1983}, just after this claim,
a shadow of doubt seems to be cast by affirming:
"This is remarkable, since the spatial velocity itself is not objective as we shall see
immediately in the proof."
\footnote{{\ }%
Objectivity means covariance by push under a space transformation.
One needs to be aware that
space projection and push by a spacetime transformation are not commutative operations.
}
Proposition \ref{prop: obj} below
provides the general result in the spacetime context
and shows that the statement in
Theorem 6.19 of \cite{MarsdenHughes1983} 
holds true only in the trivial case of a null relative space velocity between observers.
\end{remark}
\medskip

The proper treatment 
enlightens the difference between two distinct items:
\begin{itemize}
\item
A rigid movement $\,\moto_\lapse:\TjE\mapsto\TjE\,$
in the time lapse $\,\lapse\in\ZEIT\,$
with respect to a given frame which 
by preservation of simultaneity Eq.\eqref{fm: diagmotouno}
fulfils invariance of the spatial bundle
as expressed by Eq.\eqref{fm: motionspainv}.
\item
A \Euclid\ change of frame 
$\,\map:\EVE\mapsto\EVE\,$,
which by Eq.\eqref{fm: spainv}
also fulfils invariance of the spatial bundle.
Note that the requirement of isometry, proper of \Euclid\ transformations, 
is not evidenced being irrelevant.
\end{itemize}

The spacetime formulation 
depicted in the diagram of
Eq.\eqref{fm: pushmoto}
shows that the motion velocity 
$\,\Vel\in\TTjE\,$ and 
the velocity of the pushed motion
$\,\Velmap\in\TT\Tjs\,$
are related by push
according to the change of frame $\,\map:\EVE\mapsto\EVE\,$:
\begin{equation}
\setlength{\jot}{8pt}
\begin{aligned}
\Velmap&\,=\parder\lapse0\mapmoto_\lapse
\\
&\,=\TT\map\circ\,\di{\parder\lapse0\moto_\lapse}\circ\inv\map
=\map\push\Vel\,.
\end{aligned}
\label{fm: Velst}
\end{equation}

Comparing the splitting of the spacetime velocity $\,\Velmap\,$
of the pushed motion
with the push of the splitting Eq.\eqref{fm: split} of the motion velocity 
$\,\Vel=\parder{\lapse}{0}\moto_\lapse,$, we get:
\begin{equation}
\setlength{\jot}{8pt}
\left\{
\begin{aligned}
&\,\Velmap=\velmap+\timearrow\,,
\\
&\,\map\push\Vel=\map\push\vel+\map\push\timearrow\,.
\end{aligned}
\right.
\label{fm: pushvert}
\end{equation}

From Eq.\eqref{fm: Velst}-\eqref{fm: pushvert} we infer the
following basic
relation between the space velocity of the pushed motion $\,\velmap\,$,
the pushed velocity of the spatial motion $\,\map\push\vel\,$
and the spatial component 
$\,\velrel\equaldef\map\push\timearrow-\timearrow\,$
of the relative velocity of frames $\,\Velrel\equaldef\map\push\timearrow\,$:
\begin{equation}
\velmap=\map\push\vel+\velrel\,.
\label{fm: horivelpush}
\end{equation}

The computations in Eq.\eqref{fm: Velst}-\eqref{fm: pushvert}
are based on the amendment brought about by
Eqs.\eqref{fm: pushmoto}-\eqref{fm: EUtanmot}
and clarify that in the spacetime context 
\emph{covariance}
(also named \emph{objectivity} in literature
when restricted to frame changes in the \Euclid\ group)
of the spacetime velocity $\,\Vel\in\TTjE\,$
holds true under any change of observer,
while space velocities fulfils the transformation rule 
Eq.\eqref{fm: horivelpush} involving the relative space velocity.
This result is in accord with the transformation of space velocity
due to an \Euclid\ change of frame
exposed in \cite[Eq.(8), ch.VII \msec 20, p.141]{Gurtin1981}.

According to \Truesdell\ \&\ \Noll\ 
\cite[Eq.(17.3)]{TruesdellNoll1965}, 
the requirement of \emph{objectivity},
of the spatial component $\,\vel\in\HH\TjE\,$
of the velocity, consists of the transformation:
\begin{equation}
\velmap=\map\push\vel=\BQ\vel\circ\inv\map\,.
\label{fm: objspa}
\end{equation}

From Eq.\eqref{fm: horivelpush} we see
this equality is fulfilled only if 
$\,\velrel=\Bzero\,$,
i.e. there is no spatial relative velocity
between the involved frames.%
\footnote{{\ }%
The relative velocity of another observer
with respect to the privileged observer
is the push $\,\map\push\timearrow\,$
of the velocity $\,\timearrow\,$ of the spatially standing motion.
In \Newton\ group of frame-changes
Eq.\eqref{fm: newton} is fulfilled and
the spatial component of the relative velocity
is the difference $\,\map\push\timearrow-\timearrow\,$ \cite{objectivity2018}.
}

In 3D treatments reproducing the one in \cite{TruesdellNoll1965},
time is a just an ordering parameter and
the time-arrows field $\,\timearrow\,$ is not even conceived,
so that the evaluation above is out of reach.
The analysis underlines
the importance of adopting a 4D spacetime treatment of 
Continuum Mechanics (\CM).

Scholars of \CM\ must be aware that
a 3D spatial approach, with a scalar time parameter,
makes the statement of involved prioperties possibly confusing 
and the treatment prone to problematic issues. 

\goodbreak
A noteworthy instance is provided by the next statement
which, on the basis of Eq.\eqref{fm: Velst} and Eq.\eqref{fm: horivelpush},
amends Th. 6.19 of \cite{MarsdenHughes1983}
whose proof, damaged by decisive flaws,
led to the mistaken conclusion
quoted in Remark \ref{rem: MH}, affirming that
along the motion:
"The Lie-derivative of an objective tensor is objective"
\cite[end of Box 6.2, p.104]{MarsdenHughes1983}.
\medskip
\begin{proposition}[Naturality of Lie-derivatives]
\label{prop: obj}
Let $\,\moto_\lapse:\TjE\mapsto\TjE\,$ be a motion along the dynamical trajectory $\,\TjE\,$
and
$\,\map:\EVE\mapsto\EVE\,$ a change of frame. 
A tensor field $\,\BT\in\TENS\di{\TTjE}\,$ on the tangent bundle $\,\TTjE\,$
is pushed to tensor field $\,\map\push\BT\in\TENS\di{\TTjE}\,$
and the \Lie\ derivative $\,\Lieder_{\Vel}\di{\BT}\,$ 
along the spacetime motion with velocity $\,\Vel:\TjE\mapsto\TTjE\,$,
thanks to Eq.\eqref{fm: Velst} is pushed to:
\begin{equation}
\begin{aligned}
\map\push\Bigdi{\Lieder_{\Vel}\di{\BT}}
=\Lieder_{\Velmap}\di{\map\push\BT}
=\Lieder_{\map\push\Vel}\di{\map\push\BT}\,.
\end{aligned}
\label{fm: pushLiest}
\end{equation}
A tensor field $\,\BT\in\TENS\di{\HTjE}\,$ on the horizontal bundle $\,\HTjE\,$
is pushed to $\,\map\push\BT\in\TENS\di{\HTjE}\,$
and the \Lie\ derivative $\,\Lieder_{\vel}\di{\BT}\,$ along the space motion is pushed to:
\begin{equation}
\map\push\Bigdi{\Lieder_{\vel}\di{\BT}}
=\Lieder_{\map\push\vel}\di{\map\push\BT}\,,
\label{fm: pushLiehori}
\end{equation}
where, by Eq.\eqref{fm: horivelpush}:
\begin{equation}
\Lieder_{\map\push\vel}\di{\map\push\BT}
=\Lieder_{\velmap}\di{\map\push\BT}
-\Lieder_{\velrel}\di{\map\push\BT}\,,
\label{fm: Liehori}
\end{equation}
with 
$\,\vel=\Vel-\timearrow\,$
horizontal component of the velocity $\,\Vel\,$ of the spacetime motion,
$\,\Velrel=\map\push\timearrow\,$
relative velocity of the new frame with respect to the privileged one,
$\,\velrel=\Velrel-\timearrow\,$
horizontal (space) component of the relative velocity.
The last term at the r.h.s. of Eq.\eqref{fm: Liehori} amends
Th. 6.19 of \cite{MarsdenHughes1983}.
\end{proposition}

 \medskip
 \begin{remark}[Restatement in terms of flows]\label{rem: velrel}
To further clarify the difference between
Eq.\eqref{fm: Velst} and Eq.\eqref{fm: horivelpush},
we recall frame changes in the \Euclid\ group
$\,\map:\EVE\mapsto\EVE\,$
fulfil invariance of the horizontal bundle Eq.\eqref{fm: spainv}.
The relative velocity 
$\,\Velrel=\map\push\timearrow\,$
between the frames
is a spacetime vector field whose horizontal and vertical components are respectively
$\,\velrel=\map\push\timearrow-\timearrow\,$ and $\,\timearrow\,$.
In terms of flows Eq.\eqref{fm: pushLiehori} writes:
\begin{equation}
\begin{aligned}
\map\push\Bigdi{\Lieder_{\moto}\di{\BT}}
=\Lieder_{\map\push\moto}\di{\map\push\BT}
\,.
\end{aligned}
\label{fm: pushLieflow}
\end{equation}
Performing the split in Eq.\eqref{fm: motosplitframe}
and Eq.\eqref{fm: veldecomposition},
naturality of the \Lie-derivative gives:
\begin{equation}
\setlength{\jot}{8pt}
\begin{aligned}
\map\push\bigdi{\Lieder_{\moto}\di{\BT}}
&\,=\map\push\bigdi{\Lieder_{\di{\motoS\circ\motoZ}}\di{\BT}}
\\
&\,=\Lieder_{\map\push\di{\motoS\circ\motoZ}}\di{\map\push\BT}
\\
&\,=\Lieder_{\di{\map\push\motoS}\circ\di{\map\push\motoZ}}\di{\map\push\BT}
\\
&\,=\Lieder_{\map\push\motoS}\di{\map\push\BT}
+\Lieder_{\map\push\motoZ}\di{\map\push\BT}\,.
\end{aligned}
\label{fm: }
\end{equation}

The \Euclid\ frame change $\,\map\,$
fulfils Eq.\eqref{fm: spainv} so that
the push of the horizontal flow $\,\map\push\motoS\,$ is still horizontal.
On the other hand, the push of the vertical flow $\,\map\push\motoZ\,$
can be split into a commutative chain of horizontal and vertical flows:
\begin{equation}
\map\push\motoZ=\di{\map\push\moto^\VV}^\HH\circ\di{\map\push\moto^\VV}^\VV
=\di{\map\push\moto^\VV}^\VV\circ\di{\map\push\moto^\VV}^\HH\,,
\label{fm: }
\end{equation}
so that:
\begin{equation}
\setlength{\jot}{8pt}
\begin{aligned}
\Lieder_{\velmap}\di{\map\push\BT}
=\Lieder_{\map\push\motoS}\di{\map\push\BT}
+\Lieder_{\di{\map\push\moto^\VV}^\HH}\di{\map\push\BT}\,.
\end{aligned}
\label{fm: flowdec}
\end{equation}
\end{remark}
In terms of velocities, being
$\,\velrel=\parder{\lapse}{0}\di{\map\push\moto^\VV}^\HH\,$,
Eq.\eqref{fm: flowdec} writes:
\begin{equation}
\Lieder_{\velmap}\di{\map\push\BT}
=\Lieder_{\map\push\vel}\di{\map\push\BT}+\Lieder_{\velrel}\di{\map\push\BT}\,.
\label{fm: }
\end{equation}
The expression in Eq.\eqref{fm: Liehori} is thus recovered.
  

\section{MFI evaporation}
\label{sec: evaporation}

We close this observation by pointing out that
the communication between two observers is realised by means of the
pertinent diffeomorphism $\,\map:\EVE\mapsto\EVE\,$
so that any comparison should exclusively be carried out in terms of this map.

A preliminary question to be answered is the following:\\
\emph{Let a material field be detected by a privileged observer.
How does it appear to the privileged observer when evaluated by another observer?}.\\
The answer can only be in terms of push by 
the change of frame $\,\map:\EVE\mapsto\EVE\,$
\cite{objectivity2018}.

Denoting by $\,\BM\,$ the collection of material tensors 
involved in a constitutive relation according to the privileged observer,
the fields $\,\BM_\map\,$ evaluated by the new observer are got by push forward
along the transformation map:
\begin{equation}
\BM_\map\equaldef\map\push\BM\,.
\label{fm: materpush}
\end{equation}

Let us conveniently enunciate the constitutive relation by the condition:
\begin{equation}
\cR\di{\BM}=\textsc{TRUE}\,.
\label{fm: }
\end{equation}
\medskip\goodbreak

The requirement of 
\emph{Material Frame Indifference} (\MFI) claims that,
according to the privileged observer,
the constitutive relation $\,\cR_\map\,$ evaluated by the new observer 
has to fulfil the equivalence:
\begin{equation}
\cR\di{\BM}=\textsc{TRUE}\equi
\cR_\map\di{\map\push\BM}=\textsc{TRUE}\,.
\label{fm: req}
\end{equation}
It is convenient to define the push $\,\map\push\cR\,$ of the response $\,\cR\,$ by the identity:
\begin{equation}
\di{\map\push\cR}\di{\map\push\BM}
\equaldef\map\push\bigdi{\cR\di{\BM}}\,,\quad\perogni\BM\,.
\label{fm: defRpush}
\end{equation}
The indifference requirement in Eq.\eqref{fm: req} writes then as a property of covariance:
\begin{equation}
\cR_\map\equaldef\map\push\cR\,.
\label{fm: matcov}
\end{equation}

According to this analysis,
 we may conclude the so called 
\emph{Principle of Material Frame Indifference} (\MFI)
is a natural and direct consequence of the definition of
transformed response induced by a change of frame.

The  \emph{material objectivity} proposed by 
Stanis{\l}aw \Zaremba\ in 
\cite{Zaremba1903},
was reformulated by Walter \Noll\ in
\cite{Noll1955} as \emph{The Principle of Isotropy of Space} 
and renamed \emph{Material Frame Indifference} 
by Clifford \Truesdell\ in
\cite{TruesdellNoll1965}.

A lucid account by Gregory \Ryskin\   
\cite{Ryskin1985}  
stressed the necessity of a spacetime approach
to the matter, but without notable success.

The axiomatics on
Material Frame Indifference (\MFI)
has been first treated by \Marsden\ and \Hughes\ in 
\cite{MarsdenHughes1983}
in terms of \Lie-derivatives although in the unsuitable spatial context,
and repeatedly discussed by scholars 
of the \CM\ community
\cite{IbrahimbegovicTaylor2002,LiuCMT2003,LiuJEL2003,Murdoch2003,%
LiuCMT2005,MurdochCMT2005,FrewerActaMech2009,%
LiuSampaioActaMech2014,Frewer2016,LiuLee2017}.

Our proposition \ref{prop: obj}
puts in evidence the necessity of a spacetime approach
to clarify notions and treatments.
The surprising result concerning spatial objectivity
of \Lie-derivatives exposed in
\cite{MarsdenHughes1983} was shown to result from a wrong proof,
here amended by Proposition \ref{prop: obj}.

In the spacetime context,
commutativity between push by a frame transformation
and \Lie\ derivative along the spacetime motion leads readily to conclude 
that if a material tensor field is objective also its convective rate along the spacetime motion is such,
as stated in Eq.\eqref{fm: pushLiest}.

The novel geometric analysis here carried out in the spacetime context
reveals the requirement expressed by
\MFI\
is coincident with the univocal definition of modified
constitutive response due to a change of observer.
Thus the whole matter boils down to a
trivial affair \cite{Treatise2023,GCTFI2013,GNLE2014,Innsbruck2014,objectivity2018}.

\section{Integration along the trajectory}
\label{sec: Vardom}

Let $\,\conf\,$ be a body configuration, 
intersection of the spacetime trajectory with a spatial slice $\,\cS_\conf\,$,
with $\,\pconf\,$ the boundary of $\,\conf\,$.
To Carl Gustav Jacob \Jacobi\ we owe a basic formula for integrals
along the motion:
\begin{equation}
\integrale{\moto_\lapse\di\conf}{}\Boo
=\integrale{\conf}{}\di{\moto_\lapse\pull\Boo}\,.
\label{fm: jacobi}
\end{equation}

In Eq.\eqref{fm: jacobi} the symbol
$\,\Boo\in\FORME^{\nn}\di\HTj\,$
denotes any volume form on the material bundle
with $\,1\le\nn\le3\,$ geometric dimension of the body configurations.

The rate expression yields the transport formula:
\begin{equation}
\parder\lapse0\integrale{\moto_\lapse\di\conf}{}\Boo
=\integrale{\conf}{}\parder\lapse0\di{\moto_\lapse\pull\Boo}
=\integrale{\conf}{}\Lieder_{\Vel}\di\Boo\,.
\label{fm: jacobirate}
\end{equation}

This formula will be resorted to in 
\msec\ref{sec: spacetimemotion}
dealing with fundamentals of spacetime motions
and specifically in discussing about the \Euler\ and \Dalembert\
laws of dynamics Eq.\eqref{fm: EulerDalembert}.
The spatial extrusion formula \cite{Treatise2023},
with $\,\vel:=\Vel-\timearrow\,$ writes:
\begin{equation}
\integrale{\conf}{}\Lieder_{\vel}\di\Boo
=\integrale{\conf}{}\extder\di{\Boo\punto\vel}
+\integrale{\conf}{}\di{\extder\Boo}\punto\vel\,.
\label{fm: spaextr}
\end{equation}

The definition of divergence
in terms of the metric volume $\,\volformg\,$,%
\footnote{%
The metric volume is defined by the property that
a cube with unit sides has a unit volume.
}
gives the result:
\begin{equation}
\Lieder_{\vel}\di\volformg
=\der(\volformg\punto\vel)+{\di{\extder\volformg}\punto\vel}
=\diverg\di{\vel}\punto\volformg\,.
\label{fm: spaextr}
\end{equation}
Then, \Stokes-\Volterra\ integral formula Eq.\eqref{fm: Volterra} 
leads to the integral divergence theorem:
\begin{equation}
\setlength{\jot}{6pt}
\begin{aligned}
\integrale{\conf}{}\Lieder_{\vel}\di\volformg
&\,=\integrale{\conf}{}\extder(\volformg\punto\vel)
\\
&\,=\integrale{\conf}{}\diverg\di{\vel}\punto\volformg
=\ointegrale{\pconf}{}(\volformg\punto\vel)\,.
\end{aligned}
\label{fm: divTh}
\end{equation}
\medskip

\section{Laws of Motion in Spacetime}
\label{sec: spacetimemotion}

Troubles consequent to the assumption of a referential description
do dissolve \emph{ab initio}
by adopting a spacetime approach.

Indeed, constitutive relations 
are imposed on fields in the material bundle,
while equilibrium and kinematic compatibility conditions 
are imposed on fields in the spatial bundle,
as defined in \msec\ref{sec: msr}.

Constitutive  relations do involve 
material tensor fields
such as the stress and virtual stretching fields
together with their convective derivatives along the motion.

All these tensor fields do live in the material bundle
whose fibres are material body configurations,
intersections of the trajectory
with spatial slices, horizontal integral submanifolds
of the event spacetime.

A (synchronous) virtual motion from a configuration $\,\conf\,$
is a one-parameter group of virtual movements:
\begin{equation}
\vmoto_\lambda:\EUconf\mapsto\EUconf\,,
\label{fm: virtmove}
\end{equation}
i.e. automorphisms 
in the spatial slice $\,\EUconf\,$ containing $\,\conf\,$,
with $\,\vmoto_0:\conf\mapsto\conf\,$ the identity map,
as described by the commutative diagram:
\begin{equation}
\begin{aligned}
\xymatrix@R=20pt@C=60pt{
{\EUconf}
\ar[r]^{\vmoto_\lambda}
\ar[d]_{\tE}
&{\EUconf}
\ar[d]^{\tE}
\\
{\ZEIT}
\ar[r]^{\ID\ZEIT}
&{\ZEIT}
\ar[l]
}\end{aligned}
\quad\equi\quad
\tE\circ\vmoto_\lambda=\tE\,.
\label{fm: virtualmotion}
\end{equation}

The virtual velocity is the spatial field given by
$\,\dvect=\parder\lambda0\vmoto_\lambda:\conf\mapsto\TEUconf\,$.

The acceleration of the motion, 
detected in the configuration $\,\conf\,$ by an inertial observer,
is the spatial field given by the parallel derivative:
\begin{equation}
\accel
\equaldef\nabla_{\Vel}\di{\vel}\,,
\label{fm: acceldef}
\end{equation}
  
In the wake of \cite{TruesdellToupin1960},
many scholars take as definition of acceleration
the additive decomposition suggested by the
split $\,\Vel=\vel+\timearrow\,$ stated in Eq.\eqref{fm: split}:
\begin{equation}
\accel=\nabla_{\timearrow}\di{\vel}+\nabla_{\vel}\di{\vel}\,.
\label{fm: accvelsplit}
\end{equation}
This split evaluation is feasible only when 
the involved derivatives are well-defined.

This is the case at internal points of a body configuration $\,\conf\,$ 
of maximal geometric dimension 3, 
but the split cannot be applied to lower dimensional 
models of bullets, wires and membranes.

In 3D Fluid-Dynamics Eq.\eqref{fm: accvelsplit}
is the basis for \Navier-\Stokes-\StVenant\ non-linear equation of motion
for incompressible fluids.

A celebrated formula due to \Euler\ yields the material stretching field
$\,\stretching_{\dvect}\,$ of a $3D$ continuous body
undergoing a virtual motion from
a configuration $\,\conf\subset\EUconf\,$.

In \Euler\ formula, the material stretching
is expressed as half the convective (\Lie) derivative
of the spatial metric tensor 
$\,\metric:\HEVE\mapsto\HEVEs\in\COV\di\HEVE\,$
along a virtual motion
$\,\vmoto_\lambda:\EUconf\mapsto\EUconf\,$:%
\footnote{{\ }%
A basic role is played by
the \Euclid\ spatial metric tensor
$\,\metric\in\COV\di\HEVE\,$ 
and by its convective derivative
in providing an implicit definition
of rigidity constraint 
and thence in formulating the notion of equilibrium. 
Purported mathematical treatments of Mechanics in a metric-free context
\cite{Segev2023,Segev2025}
are therefore \emph{ab initio} bound to have no physical relevance.}


\begin{equation}
\setlength{\jot}{6pt}
\begin{aligned}
\stretching_{\dvect}\equaldef
\unmezzotext\,\Lieder_{\dvect}\di\metric
&\,=\unmezzotext\,\parder{\lambda}0\di{\vmoto_\lambda\pull\metric}
=\metric\punto\Bigdi{\sym\nabla\di\dvect}\,.
\end{aligned}
\label{fm: stretching3D}
\end{equation}

In Eq.\eqref{fm: stretching3D},
$\,\lambda\in\Re\,$ is a virtual-time parameter
with an arbitrary physical dimension
and the virtual-velocity field is:
\begin{equation}
\dvect=\parder{\lambda}{0}\vmoto_\lambda:\conf\mapsto\TT_\conf\EU\,.
\label{fm: }
\end{equation}

The pull-back $\,\di{\vmoto_\lambda\pull\metric}_\evento\,$
of $\,\metric\in\COV\di\HEVE\,$ at an event $\,\evento\in\conf\,$
is defined 
for $\,\Ba,\Bb\in\TT_\evento\conf\,$
in terms of the tangent functor $\,\TT\,$
by the expression:%
\begin{equation}
\di{\vmoto_\lambda\pull\metric}_\evento\di{\Ba,\Bb}\equaldef
\metric_{\vmoto_\lambda\di\evento}
\coppia{\TT_{\evento}\vmoto_\lambda\punto\Ba}{\TT_{\evento}\vmoto_\lambda\punto\Bb}\,.
\label{fm: pullback}
\end{equation}

The last equality in Eq.\eqref{fm: stretching3D} 
holds if the linear connection $\,\nabla\,$ is 
\LeviCivita, that is torsion-free and metric,%
\footnote{{\ }%
In a manifold $\,\mani\,$ endowed with a linear connection,
the vector-valued torsion two-form $\,\torsform\,$ is defined by
$\,\torsform\coppia{\Bu}{\Bv}\equaldef\nabla_\Bu\di\Bv-\nabla_\Bv\di\Bu-{\Lbrack{\Bu}{\Bv}}\,$
for all $\,\Bu,\Bv:\mani\mapsto\Tmani\,$
and $\,\Lbrack{\Bu}{\Bv}=\Lieder_{\Bu}\di{\Bv}\,$
\cite{Spivak1970,Treatise2023}.
The fundamental theorem of \Riemann\ geometry ensures existence of a unique
connection compatible with given fields $\,\nabla\metric\,$ and $\,\torsform\,$
\cite{Treatise2023}.
A linear connection is \LeviCivita\ \cite{LeviCivita1925}
if it is metric 
$\,\nabla\metric=\Bzero\,$
and torsion-free $\,\torsform=\Bzero\,$.
}
as indeed is the \Euclid\ connection
\cite[Eq.(2.10.29)]{Treatise2023}.

The formula in Eq.\eqref{fm: stretching3D}
is suitable to be applied only to interior points of continuous bodies of the 
maximal spatial geometric dimension $3D$.

In this case in fact, spatial velocities 
are also material, 
being immersions of vectors tangent to the body configuration $\,\conf\,$.

For lower dimensional continuous bodies, 
with geometric dimension 0, 1 or 2 
(bullets,wires,membranes)
it the trajectory $\,\Tj\,$
has to be taken a manifold of its own with \emph{injective immersion}%
\footnote{{\ }%
The inclusion $\,\immers:\Tj\mapsto\EVE\,$ 
is an immersion if $\,\TT\immers:\TTj\mapsto\TEVE\,$
is pointwise injective.
}
$\,\immers:\Tj\mapsto\EVE\,$ in the spacetime manifold and range 
$\,\TjE=\immers\di\Tj\,$.
\Euler's definition in Eq.\eqref{fm: stretching3D} 
must then be modified by acting with the pull-back \cite{EUL2013}:
\begin{equation}
\setlength{\jot}{6pt}
\begin{aligned}
\stretching_{\dvect}\equaldef&\,
\immers\pull\Bigdi{\unmezzotext\,\Lieder_{\dvect}\di\metric}
\\
=&\,\bigdi{\immers\pull\metric}\punto\Bigdi{\BPi\punto\Bigdi{\sym\nabla\di\dvect}\punto\BPi^A}\,.
\end{aligned}
\label{fm: eulerstretch}
\end{equation}
Here $\,\BPi^A=\TT\immers\,$
is the tangent inclusion in $3D$ space 
which is $\metric$-adjoint to the
$\,\metric$-projection operation $\,\BPi\,$
on the $1D$, $2D$ or $3D$ body configuration
\cite{Rate2014,Treatise2023}.

For $3D$ bodies the immersion and its tangent map are taken as identities.

\goodbreak

The celebrated laws conceived 
almost at the same time by
Jean-Baptiste Le Rond \Dalembert\
\cite{dAlembert1743}
and by Leonhard \Euler\
\cite{Euler1744},
for governing equilibrium in rigid body Dynamics,
are extended to deformable bodies by considering 
arbitrary (even deforming) virtual velocity fields
$\,\dvect\in\cL\,$ with $\,\cL\,$ 
linear subspace of spatial virtual velocities
conforming with firm bilateral smooth constraints.

To this end, 
the external force
due to body action $\,\Bb\,$ at distance per unit body volume $\,\volformg\,$
and to surface traction $\,\Bt\,$ by contact per unit boundary surface area $\,\pvolformg\,$:
\begin{equation}
\scalar{\fext}{\dvect}_{\conf}
\equaldef\integrale{\conf}{}\scalar{\Bb}{\dvect}\punto\volformg
+\ointegrale{\pconf}{}\scalar{\Bt}{\dvect}\punto\pvolformg
\,,
\label{fm: finn}
\end{equation}
is decomposed as sum of the internal force plus the dynamical force:
\begin{equation}
\fext\coppia{\Bb}{\Bt}\equaldef\finn\di\stress+\fdyn\,,
\label{fm: fdyn}
\end{equation}

The internal force is expressed by the principle of virtual power
with the natural stress $\,\stress\in\CON\di{\HTj}\,$
conceived as \Lagrange\ multiplier for the rigidity constraint
$\,\stretching_{\dvect}=\Bzero\in\COV\di{\HTj}\,$
on the virtual velocity:
\begin{equation}
\scalar{\finn\di\stress}{\dvect}_{\conf}
\equaldef\integrale{\conf}{}
\scalar{\stress}{\stretching_{\dvect}}\punto\massform\,.
\label{fm: finn}
\end{equation}

The natural stress $\,\stress\in\CON\di{\HTj}\,$ 
is a contravariant material tensor field
performing virtual power per unit mass by duality with the material covariant 
\Euler\ virtual stretching tensor 
$\,\stretching_{\dvect}\in\COV\di{\HTj}\,$ 
introduced in Eq.\eqref{fm: eulerstretch}.

In the literature on Continuum Mechanics, on the wake of the treatment
by Augustin-Louis \Cauchy\ 
\cite{Cauchy1829},
it is customary to consider the \emph{true} stress tensor 
$\,\BT:\HTj\mapsto\HTj\in\MIX\di{\HTj}\,$
related to the natural stress 
$\,\stress:(\HTj)^*\mapsto\HTj\in\CON\di{\HTj}\,$ 
by composition with the metric
$\,\metric:\HTj\mapsto(\HTj)^*\in\COV\di{\HTj}\,$:
\begin{equation}
\BT\equaldef\stress\punto\metric\,.
\label{fm: stress}
\end{equation}
Then, the  \emph{proof} of $\,\metric\,$-symmetry 
of $\,\BT\,$
is based on
an argument relying on the rotational equilibrium condition,
see e.g. \cite{Gurtin1972,Gurtin1981}.

The notion of stress tensor 
$\,\stress\in\CON\di{\HTj}\,$ 
as \Lagrange\ multiplier reveals
a more basic fact.

The stress tensor may in fact be assumed to be symmetric
with no loss of generality
because it is required to interact with the symmetric stretching tensor 
$\,\stretching_{\dvect}\in\COV\di{\HTj}\,$.

This is due to vanishing of interaction
between symmetric covariant tensors
and skew-symmetric contravariant tensors, 
an algebraic property in which equilibrium plays no role.

A further investigation qualifies the rotational proof of symmetry
as tautological, being based on the assumption of absence of body couples,
an assumption equivalent to symmetry \cite{Genesis2022,Treatise2023}.

The mass-form $\,\massform\,$ is of maximal order in the material bundle 
and is assumed to fulfil covariance along the motion
(a.k.a. conservation of mass):
\begin{equation}
\setlength{\jot}{6pt}
\begin{aligned}
&\,\moto_\lapse\pull\massform=\massform\
\equi
\\
&\,\Lieder_{\Vel}\di\massform=\parder\lapse0\di{\moto_\lapse\pull\massform}=\Bzero\,.
\end{aligned}
\label{fm: masscov}
\end{equation}

According to the extension of \Dalembert\ law 
to Continuum Dynamics, 
the dynamical force is expressed
in terms of the acceleration field Eq.\eqref{fm: acceldef} by
the variational equation:%
\footnote{{\ }%
In the sequel 
we will abusively identify $\,\conf\subset\Tj\,$ 
and $\,\immers\di\conf\subset\TjE\,$
to simplify notations.
}
\begin{equation}
\setlength{\jot}{6pt}
\begin{aligned}
\scalar{\fdyn}{\dvect}_{\conf}
=\integrale{\conf}{}
\di{\metric\accel\otimes\massform}\punto{\dvect}
=\integrale{\conf}{}
\metric\coppia{\accel}{\dvect}\punto\massform\,.
\end{aligned}
\label{fm: Dalembert}
\end{equation}

On the other hand,
\Euler's  law is expressed by 
stating equality between
the virtual power expended in any virtual motion by the dynamical force
and the time rate of variation of the momentum projected on the virtual velocity:
\begin{equation}
\setlength{\jot}{6pt}
\begin{aligned}
\scalar{\fdyn}{\dvect}_{\conf}
&\,=\parder\lapse0\integrale{\moto_\lapse\di\conf}{}
\Bigdi{\metric\vel\otimes\massform}\punto\dvect
\\
&\,=\parder\lapse0\integrale{\moto_\lapse\di\conf}{}
\metric\coppia{\vel}{\dvect}\punto\massform
\,.
\end{aligned}
\label{fm: Eulerdyn}
\end{equation}

Equilibrium and kinematic compatibility are thus expressed in terms of 
spatial tensor fields based on the trajectory but living in the spatial bundle
whose fibres are spatial slices.

A conforming virtual velocity $\,\dvect:\conf\mapsto\TT_{\conf}\EU\,$
is a smooth spatial tangent vector field fulfilling the imposed linear kinematical constraints
prolonged along the motion by parallel transport.
Denoting by $\,\nabla\,$ the \Euclid\ connection 
associated with the parallel transport by translation,
we have:
\begin{equation}
\nabla_{\Vel}\di\dvect=\Bzero\,.
\label{fm: paraltrans}
\end{equation}
The importance of this prolongation will appear in 
Eq.\eqref{fm: EulerDalembert}.

The spatial metric $\,\metric\,$ is uniform in \Euclid\ spacetime:
\begin{equation}
\nabla\di\metric=\Bzero\,.
\label{fm: metricconn}
\end{equation}

Let us now prove equivalence between
the law of motion due to \Dalembert\
\cite{dAlembert1743},
expressed in terms of acceleration field,
and the one due to \Euler\ 
\cite{Euler1744},
formulated in terms of rate of change of the kinematical momentum.

To this end, assume both Eq.\eqref{fm: masscov}
expressing conservation of mass along the motion,
and Eq.\eqref{fm: metricconn}
expressing invariance of the metric under parallel transport:
\begin{equation}
\setlength{\jot}{8pt}
\left\{
\begin{aligned}
&\,\Lieder_{\Vel}\di\massform=\Bzero\,,
\\
&\,\nabla\di\metric=\Bzero\,.
\end{aligned}
\right.
\label{fm: twocond}
\end{equation}

Let us recall that
$\,\conf\,$ is the body spatial configuration,
$\,\Vel:\conf\mapsto\TT_{\conf}\TjE\,$  is the spacetime velocity field,
and $\,\vel:\conf\mapsto\TT_{\conf}\EU,$ is the spatial component.

Equivalence between \Dalembert\ and \Euler\ laws
is then proven as follows:
\begin{equation}
\setlength{\jot}{12pt}
\begin{aligned}
\scalar{\fdyn}{\dvect}_{\conf}
&\,=\parder\lapse0\integrale{\moto_\lapse\di\conf}{}
\Bigdi{\metric\vel\otimes\massform}\punto\dvect
=\parder\lapse0\integrale{\moto_\lapse\di\conf}{}
\metric\coppia{\vel}{\dvect}\punto\massform
\\
&\,=\integrale{\conf}{}\parder\lapse0\moto_\lapse\pull\Bigdi{
\metric\coppia{\vel}{\dvect}\punto\massform}
=\integrale{\conf}{}\Lieder_{\Vel}\Bigdi{
\metric\coppia{\vel}{\dvect}}\punto\massform
\\
&\,=\integrale{\conf}{}
\nabla_{\Vel}\Bigdi{\metric\coppia{\vel}{\dvect}}\punto\massform
=\integrale{\conf}{}\metric\coppia{\accel}{\dvect}\punto\massform
\,.
\end{aligned}
\label{fm: EulerDalembert}
\end{equation}

.

\section{Traveling control windows}
\label{sec: windows}

Problems of Dynamics are often conveniently formulated and answered
in terms of the basic laws as seen by an observer through a 
control window traveling in the dynamical trajectory
\cite{ControlWindows2018}.

Let $\,\volform:\TjE\mapsto\VOL\di\VEVE\,$
be a spatial maximal form over the trajectory $\,\TjE\,$.

We consider a control window 
$\,\control\subset\TjE\,$,
an \emph{outer}-oriented spatial manifold 
undergoing, along its own trajectory $\,\TjC\subset\TjE\,$,
a travel $\,\travel_\lapse:\TjC\mapsto\TjC\,$ such that:%
\begin{equation}
\travel_\lapse\di\control\subset\moto_\lapse\di\control\,.
\label{fm: }
\end{equation}

We may then evaluate the gap between the rates of variation of the 
$\,\volform$-volume of the control window, 
respectively evaluated:
\begin{itemize}
\item
along the travel
$\,\travel_\lapse:\TjC\mapsto\TjC\,$
with velocity
\begin{equation}
\Veltravel=\parder\lapse0\travel_\lapse:\TjC\mapsto\TT\TjC\,,
\label{fm: }
\end{equation}
\item
and along the motion
$\,\moto_\lapse:\TjE\mapsto\TjE\,$
with velocity
\begin{equation}
\Vel=\parder\lapse0\moto_\lapse:\TjE\mapsto\TTjE\,.
\label{fm: }
\end{equation}
\end{itemize}
By a geometric analysis based on Eq.\eqref{fm: divTh},
this gap is revealed to be given by 
the $\,\volform$-volumic flux of the relative spatial velocity $\,\veltravel-\vel\,$
through the boundary $\,\pcontrol\,$ of the control window $\,\control\,$:
\begin{equation}
\begin{aligned}
\parder\lapse0\integrale{\travel_\lapse\di\control}{}\volform
-\parder\lapse0\integrale{\moto_\lapse\di\control}{}\volform
&\,=\ointegrale{\pcontrol}{}\volform\punto(\veltravel-\vel)\,.
\end{aligned}
\label{fm: windowcontrol}
\end{equation}

By virtue of Eq.\eqref{fm: windowcontrol}, 
setting $\,\control=\conf\,$ and
$\,\volform=\metric\coppia{\vel}{\dvect}\punto\massform\,$
on $\,\moto_\lapse\di\conf\,$,
the \Euler\ law of motion 
Eq.\eqref{fm: Eulerdyn} of a massive body  
may be written in terms of a control window $\,\conf\,$ 
traveling along the trajectory with velocity $\,\veltravel\,$
as \cite{ControlWindows2018}:
\begin{equation}
\setlength{\jot}{6pt}
\begin{aligned}
\scalar{\fdyn}{\dvect}_\conf
&\,=\parder\lapse0\integrale{\moto_\lapse\di\conf}{}
\metric\coppia{\vel}{\dvect}\punto\massform
\\
&\,=\parder\lapse0\integrale{\travel_\lapse\di\conf}{}
\metric\coppia{\vel}{\dvect}\punto\massform
\\&\,\quad
-\ointegrale{\pconf}{}
\metric\coppia{\vel}{\dvect}\punto\massform\punto(\veltravel-\vel)\,.
\end{aligned}
\label{fm: controldyn}
\end{equation}

In Eq.\eqref{fm: controldyn},
the manifold chain $\,\control\,$ and its boundary $\,\pcontrol\,$
are assumed to be \emph{outer} oriented with compatible orientations
\cite{Advancements2024}.

In \cite[\msec{15}]{Gurtin1981} the formula Eq.\eqref{fm: controldyn}
is confined to the case of a control window fixed in space,
$\,\veltravel=\Bzero\,$ 
and therefore cannot be representative of a force.

The expression in Eq.\eqref{fm: controldyn} 
provides a direct interpretation of well-known methods of
Computational Dynamics:
\begin{itemize}
\item[1)]
The \Lagrange\ point of view 
assumes a control window in the trajectory with 
a travel spatial velocity equal to the spatial velocity of the motion $\,\veltravel=\vel\,$.
\item[2)]
The \Euler\ point of view 
assumes a control window in the trajectory with
a vanishing spatial travel velocity $\,\veltravel=\Bzero\,$.
This point of view is valid until the control window fixed in space remains 
included in the dynamical trajectory.
\item[3)]
The Arbitrary \Lagrange-\Euler\ (\ALE)
point of view assumes a control window traveling in the trajectory with
any spatial velocity $\,\veltravel:\conf\mapsto\TT_\conf\EU,$
such that the control window remains 
included in the dynamical trajectory.
\end{itemize}

The expression in Eq.\eqref{fm: controldyn} 
was resorted to in 
the analysis of motions involving the interaction of a fluid with a solid case
performed in \cite{SFCM2017}.
The treatment provides a Continuum Mechanics basis  
for the evaluation of the thrust exerted by the fluid on the solid.

The dynamics of rigid bodies with variable mass initiated 
in the realm of Analytical Mechanics by
\Buquoy
\cite{Buquoy1812,Buquoy1815}
and \Meshchersky\
\cite{Meshchersky1897,Meshchersky1949}%
\footnote{
Georg Franz August de Longueval, Baron von Vaux, Graf von Buquoy (1781--1851),\\
Ivan Vsevolodovich Meshchersky (1859 - 1935).
}
leaved opened the question 
of whether \Dalembert\ or \Euler\ law is to be applied since
the equivalence stated in Eq.\eqref{fm: EulerDalembert}
does not hold, being the mass not conserved.

Under suitable assumptions,
the thrust exerted by the fluid on the solid case
was evaluated in  \cite{SFCM2017}
without considering bodies of variable mass.

\medskip\goodbreak
The thrust $\,\Fthrust\,$
is estimated to be the loss-rate in kinetic momentum 
of the fluid in relative motion with respect to 
a control window attached to the solid case 
(the \emph{skeleton}): 
\begin{equation}
\scalar{\Fthrust}{\dvect}_\conf\equaldef
-\ointegrale{\pconf}{}
\metric\coppia{\velrel}{\dvect}\punto(\massform_\flu\punto\velrel)\,.
\label{fm: thrust}
\end{equation}

In Eq.\eqref{fm: thrust}
$\,\massform_\flu\,$ is the mass-form of the fluid
with the mass-loss $\,\massform_\flu\punto\velrel\,$ vanishing on 
the control window boundary where it is
attached to the solid case.
The thrusting effect is due to the fluid velocity
$\,\velrel=\velflu-\velske\,$
relative to the skeleton.

\section{Reference placements and potato-shaped avatars}
\label{sec: reference}

Books and papers on fundamentals of \CM\
are often illustrated with arrows depicting maps 
between potato-shaped \emph{avatars}
pretending to provide a geometric picture of the material body 
and of its spatial placements.

This seemingly friendly manner of illustrating placements 
in the 3D ambient space has several drawbacks 
and is by no means supported by
physical insight and engineering motivation.

Many good reasons lead us to consider the 4D spacetime manifold of events 
as the stage where actions take place and to put at the centre of the scene
the trajectory manifold and the motion progressing in it
as a one-parameter group of automorphic movements
\cite{Treatise2023}.

Abstract as it may appear, this geometric construction is by far
more realistic and useful than one might think at first glance.

The potatoes point of view comes indeed readily to face with questions which are hard
to be answered in a physically meaningful manner.

A first basic question is about how to choose a diffeomorphic correspondence 
between a reference \emph{potato-shaped avatar} and the actual placement of the body. 

In fact, there is a plethora of candidate maps sharing 
the same domain and range and any chosen correspondence 
should be shown to be not influent on the physical description of the relevant phenomena.

This is a necessary but very challenging, if not impossible, mission.

Note that in Computational Mechanics practice,
such as in Finite Element Method (\FEM),  
local correspondences are constructed between 
polyhedral or simplex elements in the actual placement
and their referential counterparts,
but only piecewise in the material bundle.  

Therefore this construction could perhaps be useful for describing constitutive relations
but certainly not apt to deal with global equilibrium in the spatial bundle.

Another drawback resident in the potatoes point of view consists of
the formulation of boundary conditions for constrained continua.

The standard procedure is based on splitting the boundary
into disjoint complementary parts where
respectively static and kinematic data are imposed.

This picture, appears to have been first suggested by the mathematicians
Jacques-Louis \Lions\ and Georges \Duvaut\ in \cite{LionsDuvaut1972},
by separating the loci on the boundary surface where
static and kinematic boundary conditions,
espectively investigated by 
Carl Gottfried \Neumann\ and Johann Peter Gustav Lejeune \Dirichlet,
are respectively resident 

However difficulties are faced in the mathematical qualification
of the involved fields at the interfaces between complementary parts,
as evidenced by Franco \Brezzi\ and Michel \Fortin\ in 
\cite{BrezziFortin1991}.

Moreover, this standard procedure should profitably 
be abandoned being needlessly restrictive 
even for usual engineering modelling. 

In fact, by adopting a variational formulation,
it is readily evidenced static conditions 
may be assigned anywhere on the boundary, 
independently of imposed kinematic conditions
\cite{Treatise2023}.

This is a common good practice in engineering structural schemes.

\section{Finite Elasticity and Anelasticity}
\label{sec: Elas}

Elasticity is the basic \CM\ model
for constitutive behaviour of materials.

Classical treatments consider only linearised approximations
in which configuration changes during the motion are assumed to be negligible
in imposing equilibrium conditions.
Most powerful results and methods of analysis of elastic structures have been 
in fact developed in this simplified context.

A noteworthy exception was Leonhard \Euler\ elastic bifurcation theory
\cite{Euler1744}.

The \emph{hypo-elastic} rate model 
was proposed by \Truesdell\ \cite{Truesdell1955}
some seventy years ago as a constitutive model suitable
to describe conservative elasticity in terms of the stress and its 
\emph{corotational} time-rate.

In this respect we may quote the following comment by 
Rodney \Hill\ in \cite{Hill1959}:

\emph{%
\Truesdell\ in \cite{Truesdell1955} has isolated for special study solids for which the functions f
are linear in strain-rate and depend in addition only on the final stress, and has
proposed the name ‘ hypo-elastic.’ 
\Truesdell's intention was to formulate a new
concept of elastic behaviour or, more precisely, 
a concept of elastic behaviour expressed entirely in terms of rates. 
However, probably
the overwhelming majority of hypo-elastic solids are inelastic, the stress being
recovered only on special circuits (such as the degenerate kind mentioned earlier).
And, in general, it is not even possible to regard non-integrable rate equations as
approximately equivalent to incremental relations in a small enough neighbourhood,
since differences may not remain negligible when circuits are continually repeated.}

The hypo-elastic model was asserted to be bound to failure 
by Barry \Bernstein\ in 
\cite{BernsteinB1960},
and this conclusion was taken for granted afterwords,
until a brand new rate-theory was proposed in \cite{GNLE2014,Innsbruck2014,Treatise2023},
as summarised below in \msec\ref{sec: RateElas}.

The negative conclusion about integrability
of the hypo-elastic rate model
led Erastus Henry \Lee,
just a few years after publication of the \emph{monsterino}
\cite{TruesdellNoll1965},
to embrace in \cite{LeeEH1969}
the diabolic suggestion of treatments in reference placements.

The resulting constitutive scheme consisted of splitting
the so-called \emph{deformation gradient} $\,\BF\,$, 
improperly taken as measure of finite distortion,
into a \emph{chain} of subsequent
plastic $\,\BF_{p}\,$ and elastic $\,\BF_{e}\,$ linear transformations between
local configurations, respectively labeled 
as initial $\,\conf\,$, intermediate and current:
\begin{equation}
\BF=\BF_{e}\BF_{p}\,.
\label{fm: Bilby}
\end{equation}
This decomposition was first proposed 
in \cite{Bilby1956}
by Bruce Alexander \Bilby\  et al., as remarked in \cite{SadikYavari}.

Starting from the first proposals dating about 40 years ago
\cite{Kondaurov1987,TakamizawaHayashi1987}
and the later one in \cite{Rodriguez1994},
the chain decomposition 
has found application also in Biomechanics
as a means to split elastic and growth \&\ remodelling phenomena 
(G\&R) in biological tissues,
by changing the notation in Eq.\eqref{fm: Bilby} to:
\begin{equation}
\BF=\BF_{e}\BF_{g}\,.
\label{fm: Kondaurov}
\end{equation}

Nowadays, the list of contributions to mechanics of biomaterials
based on the multiplicative split is so large to discourage any attempt
of reproduction here,
referring to overview works on the topic, as quoted e.g. in 
\cite{Goriely2017,ZhuanLuo2022,ChenYC2024}.
 
The symbol $\,\BF\,$ in Eq.\eqref{fm: Bilby} and Eq.\eqref{fm: Kondaurov}
is a shorthand for the \emph{tangent movement}:
\footnote{{\ }%
Introduced in \cite{TruesdellNoll1965}
as \emph{deformation gradient} this notion 
is nowadays ubiquitous in pertinent literature.
However $\,\TT\moto_\lapse\,$
is not a \emph{gradient} 
but a \emph{tangent map}
and the movement $\,\moto_\lapse\,$ 
is not a \emph{deformation}.
It is  advisable to revise both the name and the notation of $\,\BF\,$
for the sake of clarity and mechanical consistency.
The ambiguous notation has been source of serious conceptual difficulties
hidden behind a crowd of algebraic expressions without mechanical relevance.
}
\begin{equation}
\BF_\lapse\equaldef\TT\moto_\lapse
:\Tconf\mapsto\TT\di{\moto_\lapse\di\conf}\,,
\label{fm: tanmove}
\end{equation}
fulfilling the commutative diagram:
\begin{equation}
\begin{aligned}
\xymatrix@R=20pt@C=60pt{
{\Tconf}
\ar[r]^{\kern-15pt\BF_\lapse\equaldef\TT\moto_\lapse}
\ar[d]_{\proj}
&{\TT\di{\moto_\lapse\di\conf}}
\ar[d]^{\proj}
\\
{\conf}
\ar[r]^{\moto_\lapse}
&{\moto_\lapse\di\conf}
}\end{aligned}
\quad\equi\quad
\proj\circ\TT\moto_\lapse=\moto_\lapse\circ\proj\,,
\label{fm: gradmotion}
\end{equation}
in which $\,\proj:\Tconf\mapsto\conf\,$ is tangent bundle projection.

The finite constitutive law according to the multiplicative (chain)
decomposition of the deformation gradient
was adopted also by Juan Carlos \Simo\ in \cite{Simo1988}
and from there on spread out in the literature
pertinent to elasto-plasticity.

This constitutive scheme is however untenable for several good reasons
\cite{Treatise2023}.

It is in manifest disagreement with the physical outcome
of mechanical experiments according to which elastic materials 
do not react to rigid transformations
which do not modify lengths of material lines in the body.

In trying to overcome the issue, 
a correction was proposed in \cite{Gurtin1981}
by a reduction argument 
to drop off the polar decomposition 
$\,\BF=\BR\BU\,$
\footnote{{\ }%
In full notation we have: $\,\BF_\lapse=\BR_\lapse\BU_\lapse\,$
with $\,\BR_\lapse:\Tconf\mapsto\TT\di{\moto_\lapse\di\conf}\,$
and $\,\BU_\lapse:\Tconf\mapsto\Tconf\,$
with $\,\BU_\lapse^2\equaldef\BF_\lapse^A\BF_\lapse\,$.
The isometry $\,\BR_\lapse\,$
fulfils $\,\metric\coppia{\BR_\lapse\Bh}{\BR_\lapse\Bh}=\metric\coppia{\Bh}{\Bh}\,$
for any $\,\Bh\in\Tconf\,$, being $\,\metric\,$ invariant under \Euclid\ distant parallel transport
\cite{Treatise2023}.
}
the rotational part $\,\BR\,$.

The procedure was however based on 
the improper formulation of \MFI\ here described in 
Eq.\eqref{fm: tanmoveTN}.
But, \emph{two wrongs do not make a right!}

Indeed, a constitutive law involving a relation between 
a state variable (the stress state)
pertaining to a single configuration
and a finite stretch depending on two configurations,
is mathematically and physically untenable.

Moreover, the chain decomposition
requires a sequential ordering of phenomena
described by non-commuting deformation gradients, with no physical motivation.

A further, more subtle difficulty lies in the very definition of the 
time derivative of the tangent motion Eq.\eqref{fm: tanmove}:
\begin{equation}
\BL=\dot\BF\equaldef
\parder\lapse0\back_\lapse(\TT\moto_\lapse)
:\Tconf\mapsto\TT_\conf\EU\,.
\label{fm: dotF}
\end{equation}

To provide stretching and spin tensorial measures.
this spatial tensor is split into symmetric and antisymmetric parts:
\begin{equation}
\BL=\BD+\BW\,.
\label{fm: }
\end{equation}

The evaluation of $\,\BL=\dot\BF\,$ in Eq.\eqref{fm: dotF}
depends on the assumed parallel transport
and leads to a spatial tensor not suitable to appear in constitutive relations,
where only material tensors are admitted to enter.

The troublesome time-rate $\,\dot\BF\,$ and its split can however be
conveniently by-passed.


A clearer insight is got
by a geometric argument 
consisting of the decomposition in symmetric and skew-symmetric parts
of the parallel derivative of the spatial motion velocity, evaluated
according to the (unique) \LeviCivita\ connection $\,\nabla\,$
associated with the spatial metric tensor field
$\,\metric:\TT_\conf\EU\mapsto\TT^*_\conf\EU\,$:
\begin{equation}
2\,\metric\punto\nabla\di{\vel}=\Lieder_{\vel}\di\metric+\extder(\metric\punto\vel)\,.
\label{fm: symskewparts}
\end{equation}

Here $\,\extder(\metric\punto\vel)\,$ is the differential two-form 
defined by the exterior derivative $\,\extder\,$ of the one-form
$\,\metric\punto\vel:\conf\mapsto\TT^*_\conf\EU\,$,
resulting from contracting the metric two-tensor:
\begin{equation}
\metric:\TT_\conf\EU\otimes_\conf\TT_\conf\EU\mapsto\FUN\di{\TT_\conf\EU}\,.
\label{fm: }
\end{equation}
with the space motion velocity $\,\vel:\conf\mapsto\TT_\conf\EU\,$
 \cite[III \msec2.10.8]{Treatise2023}, \cite{GCM2014}.

From Eq.\eqref{fm: symskewparts},
the expression of stretching and spin 
in terms of the \LeviCivita\ connection $\,\nabla\,$ are given by:
\begin{equation}
\setlength{\jot}{8pt}
\begin{aligned}
\unmezzotext\Lieder_{\vel}\di\metric&=\sym\di{\metric\punto\nabla\di{\vel}}\,,
\\
\unmezzotext\extder(\metric\punto\vel)&=\anti\di{\metric\punto\nabla\di{\vel}}\,.
\end{aligned}
\label{fm: }
\end{equation}
 
With the chain constitutive scheme,
the diabolic suggestion of reference placements
reached an apotheosis,
since a sequel of intermediate local configurations
is to be involved, in addition to a reference one.

Despite evident drawbacks, such as
unphysical occurrence of \emph{plastic spin} 
and logical bugs related to ordering of non-commutative contributions,
the proposal was labeled as \emph{multiplicative (chain) decomposition}
and, in the absence of alternative wayout
gained an undeserved success
being still adopted in elastoplasticity,
as documented by papers and books
of Vlado \Lubarda\ \cite{Lubarda1991,Lubarda2002},
\&\ Robert J. \Asaro\ \cite{AsaroLubarda2006}.%
\footnote{{\ }%
At the COMPLAS VIII 2005 meeting in Barcelona,
the first author Giovanni \Romano\
(engineer and professor of Structural Mechanics)
having attended a general lecture by Klaus-J{\"u}rgen \Bathe\
on nonlinear plasticity, asked the speaker why
not to give up with the troublesome decomposition
$\,\BF=\BF_{e}\BF_{p}\,$ after so many evidenced drawbacks.
The puzzling answer was: \emph{But we are engineers}.
}

The problematic state of the art in elasto-plasticity theory is  
well-documented in recent valuable contributions
by Otto Timme \Bruhns\
\cite{Bruhns2014,Bruhns2021}
including historical notes and a comparison between the Heinrich \Hencky\ 
finite incremental plasticity theory based on the logarithmic strain
$\,\log\di{\BU_\lapse}\,$
and the plastic flow theory of Ludwig \Prandtl\ and András (Endre) \Reuss.

The author of \cite{Bruhns2021} was however not aware 
of the geometrical advances exposed in 
\cite{Treatise2023,GCM2014,objectivity2018,GNLE2014,Advancements2024}.

On \Prandtl-\Reuss\ plastic flow theory, 
\Bruhns\ comment in \cite{Bruhns2021} was:

\emph{First problems emerged, when after World War II these relations 
were transferred to application within large deformations. 
The objectivity of the incorporated rates was questioned. 
As a consequence, several possible objective rates were discussed, 
some of them producing spurious effects in the results of calculation. 
Moreover, the (alleged) dissipative character of the elastic-like (hypoelastic) part 
of the \Prandtl-\Reuss\ equations was taken as argument to discredit these relations 
as only applicable for small – at least small elastic – deformations.}

In addition we add the remark that
a geometric approach in the context of \Euclid\ spacetime
provides a clear view of the matter and a unique answer to the question
of how to define the stress rate.
This decisive advancement is exposed in the next section.

\section{Rate Elasticity and Anelasticity}
\label{sec: RateElas}

Having become aware of the impracticability of finite formulations,
after a detailed investigation of the involved issues,
the authors eventually succeeded in conceiving and
constructing a natural formulation of nonlinear elasticity
\cite{GNLE2014,Treatise2023}.

The result is a rate model in which the outcome of the constitutive law is the
elastic-stretching:
\begin{equation}
\ela:{\Tconf\mapsto(\Tconf)}^*\in\COV\di\HTj\,.
\label{fm: }
\end{equation}

This is a symmetric covariant tensor field
linearly related to the contravariant \emph{stressing}
$\,\stressing=\Lieder_{\Vel}\di\stress\,$,
given by the \Lie\ (or convective) derivative along the motion of the 
contravariant natural stress-state 
\begin{equation}
\stress:{(\Tconf)}^*\mapsto\Tconf\in\CON\di\HTj\,,
\label{fm: }
\end{equation}
through the tangent elastic compliance $\,\BH\di{\stress}\,$,
in turn nonlinearly depending on the stress-state 
\cite{GNLE2014,Treatise2023}:
\begin{equation}
\ela\equaldef\BH\di{\stress}\punto\stressing\,.
\label{fm: rateelast}
\end{equation}
\medskip\goodbreak

The tangent elastic compliance:
\begin{equation}
\BH\di\stress:\CON\di\HTj\mapsto\COV\di\HTj\,,
\label{fm: tanelascompl}
\end{equation}
is a fiberwise linear isomorphism over the identity, 
from the contravariant stress bundle
to the covariant stretching bundle,
nonlinearly dependent on the stress state,
as shown by the commutative diagram: %
\footnote{{\ }%
The projections $\,\proj_\con\,$ and $\,\proj_\cov\,$
map the stress and stretching tensors onto their base point in the spacetime trajectory $\,\Tj\,$,
and define the stress and stretching bundles.}
\begin{equation}
\begin{aligned}
\xymatrix@R=20pt@C=40pt{
{\CON\di\HTj}
\ar[r]^{\BH\di\stress}
\ar[d]_{\proj_\con}
&{\COV\di\HTj}
\ar[d]^{\proj_\cov}
\\
{\Tj}
\ar[r]^{\ID\Tj}
&{\Tj}
}
\end{aligned}
\quad\equi\quad
\proj_{\cov}\circ\BH\di\stress=\proj_{\con}\,.
\label{fm: matmorlocC}
\end{equation}

The success of this rate theory of elasticity is due to the self-proposing choice
of the \emph{natural stress} tensor  
$\,\stress\in\CON\di\HTj\,$ 
as contravariant stress state performing
elastic power per unit mass by the duality pairing
$\,\scalar{\stress}{\stretching_\el}\,$
defined as linear invariant of the operator 
$\,\stress\circ\stretching_\el:\Tconf\mapsto\Tconf\in\MIX\di{\Tconf}\,$
post-composition of the natural stress with the covariant elastic-stretching 
$\,\ela\in\COV\di\HTj\,$,
in analogy with Eq.\eqref{fm: finn}.

Mass conservation along the body motion,
plays a basic role in setting up the classical dynamical theory,
as shown in Eq.\eqref{fm: EulerDalembert},
and is also decisive in ensuring the non-dissipative character 
of large elastic movements.

The basic elastic property 
of null energy dissipation (or production)
in \emph{material cycles} of stress-states,
made precise in Def.\ref{def: cycles},
is assured by Prop.\ref{prop: estatepot}.

Elasticity is then characterised by the property that in each configuration
the elastic compliance
$\,\BH=\derdueV\di\Gelast\,$
is the second derivative at fixed time
%
\footnote{ {\ }%
The fiber-derivative $\,\derV\,$
in the pertinent spatial slice of the 
potential $\,\Gelast\,$, is analogous to the one introduced by George \Green\ in
\cite{Green1839}, but a function of the natural stress \cite{Treatise2023}.}
of a convex stress potential 
$\,\Gelast:\CON\di{\HTj}\mapsto\FUN\di{\HTj}\,$
which is time-invariant, i.e. such that:
\begin{equation}
\dot\Gelast\equaldef\Lieder_{\Vel}\di{\Gelast}=\Bzero\,.
\label{fm: timeinvpot}
\end{equation}
\begin{definition}[Elastic states]
The elastic-state $\,\es\in\COV\di\Tconf\,$ 
is a symmetric covariant material tensor
output of the invertible nonlinear constitutive relation:
\begin{equation}
 \es=\Celast\di\stress\equaldef\derV\Gelast\di\stress\,.
\label{fm: elasticstate}
\end{equation}
\end{definition}

Taking the convective time-derivative, and applying \Leibniz\ rule we infer:
\begin{equation}
\vcenter{\halign{
\hfil$#$&$#$\hfil&$#$\hfil&$#$\hfil\cr
\des&\,\equaldef\Lieder_{\Vel}\di\es
=\Lieder_{\Vel}(\Celast\circ\stress)
\vspace{12pt}\cr
&\,=\Lieder_{\Vel}\di\Celast\circ\stress+\derV\Celast\di\stress\punto\stressing
\vspace{12pt}\cr
&\,=\dot\Celast\circ\stress+\BH\di{\stress}\punto\stressing
\vspace{12pt}\cr&\,
=(\derV\dot\Gelast)\circ\stress+\derdueV\Gelast\di{\stress}\punto\stressing
\,.\cr}}
\label{fm: stretchelastate}
\end{equation}

Time-invariance along the motion Eq.\eqref{fm: timeinvpot}, ensures that:
\begin{equation}
\begin{aligned}
\dot\Celast\equaldef
&\,\Lieder_{\Vel}\di\Celast
=\Lieder_{\Vel}\di{\derV\Gelast}
=\,\derV\Bigdi{\Lieder_{\Vel}\di{\Gelast}}
=\,\derV\dot\Gelast=\Bzero\,.
\end{aligned}
\label{fm: ratepot}
\end{equation}
and hence Eq.\eqref{fm: stretchelastate}
yields the expression of the elastic-stretching 
as \Lie-derivative of the elastic-state along the motion:
\begin{equation}
\stretching_\el=\des\equaldef\Lieder_{\Vel}\di\es\,.
\label{fm: elstaterate}
\end{equation}

The elastic relation between the
\emph{stress potential} $\,\Gelast\,$ 
and the conjugate \emph{elastic-state potential} $\,\Gelasts\,$
is provided by the \Euler-\Legendre\ transform:
\begin{equation}
\left\{
\begin{aligned}
&\,\es=\Celast\di\stress=\derV\Gelast\di\stress\,,
\\
&\,\Gelasts\di\elstate\equaldef\scalar{\stress}{\es}-\Gelast\di\stress\,,
\\
&\,\stress=\inv\Celast\di\es=\derV\Gelasts\di\es\,.
\end{aligned}
\right.
\label{fm: Legendre}
\end{equation}

A careful geometric integration of the power
performed by the stress-state on the elastic-stretching,
shows the elastic work expended
in closed cycles of natural stress
is vanishing,
a basic result formulated in Proposition \ref{prop: estatepot}
proven in 
\cite{Treatise2023,Innsbruck2014}.

Time invariance of the elastic response 
Eq.\eqref{fm: ratepot} gives
$\,\moto_\lapse\pull\Celast=\Celast\,$.

Then, covariance of the stress-state along the motion
$\,\moto_\lapse\pull\stress=\stress\,$
implies covariance of the elastic-state along the motion:
\begin{equation}
\setlength{\jot}{8pt}
\begin{aligned}
\moto_\lapse\pull\es
&\,=\moto_\lapse\pull\Bigdi{\Celast\di\stress}
\\
&\,=(\moto_\lapse\pull\Celast)\di{\moto_\lapse\pull\stress}
=\Celast\di\stress
=\es\,,
\end{aligned}
\label{fm: equiv}
\end{equation}
and vice versa by invertibility of the elastic response $\,\Celast\,$.
\medskip
\begin{proposition}[Mechanical work expended]
\label{prop: estatepot}
The internal mechanical work expended
in an elastic process of duration $\,\Delta\ttt\,$
along the dynamical trajectory
with a movement $\, \moto_{\Delta\ttt}\,$
from the configuration $\,\conf\,$ to $\,\moto_{\Delta\ttt}\di\conf\,$,
is expressed by:
\begin{equation}
\integrale{0}{\Delta\ttt}\inde\lapse
\integrale{\moto_\lapse\di\conf}{}\scalar{\stress}{\ela}\punto\massform
=\Gelasts_\conf\di{\moto_{\Delta\ttt}\pull\es}-\Gelasts_\conf\di\es\,,
\label{fm: mechintwork}
\end{equation}
where 
$\,\Gelasts_\conf\,$ is the global elastic-state potential:%
\begin{equation}
\Gelasts_\conf\di\es\equaldef\integrale{\conf}{}\Gelasts\di\es\punto\massform\,.
\label{fm: globpot}
\end{equation}
\end{proposition}

\begin{definition}[Cyclic process]
\label{def: cycles}
A material process is a movement of finite duration
involving material tensor fields along the trajectory.
A cyclic material process (or \emph{material cycle}) is such that
each involved field takes in the final configuration a value which
is the push-forward along the movement
of the field-value in the starting configuration.
\end{definition}

By this definition and Eq.\eqref{fm: equiv},
from Proposition \ref{prop: estatepot}
it readily follows that the global internal mechanical work
vanishes when the material process
is a cycle of stress-states, or equivalently a cycle of elastic-states.

Relying on the notion of elastic-state and on conservation of mass,
the statement in Prop.\ref{prop: estatepot} provides also an answer to what
\Truesdell\ called 
\emph{Das ungelöste Hauptproblem der endlichen Elastizitätstheorie}
\di{unsolved Main Problem of Finite Elasticity Theory}
\cite{Truesdell1956,Carroll2009}.

The negative conclusion about integrability of the
hypo-elasticity law exposed by \Bernstein\ in \cite{BernsteinB1960}
has thus been overcome by the rational geometric scheme
of rate-elasticity in terms of elastic states, leading to Eq.\eqref{fm: mechintwork}.

In more general situations, which are of the greatest interest for application of Continuum Mechanics
to structural engineering,
the geometric stretching is composed by additioning 
elastic stretching and non-elastic stretching,
in perfect accord with the modelling adopted by
pioneers of the small displacement elasto-plasticity theory, see e.g. \cite{Bruhns2021},
but replacing the partial time-derivative with the \Lie-derivative along the motion.

The rate model so formulated is able to take into account possibly dissipative contributions
due to visco-plastic effects, changes of temperature and internal structure,
actions of electromagnetic fields, and so on.

\section{Continua with microstructure: the issue of redundancy}
\label{sec: Microstr}

The proposal of endowing the continuum model with an
additional microstructure
is usually attributed to Eugène and Fran{\c c}ois \Cosserat\
in the treatise on the theory of deformable  bodies
drawn up at the beginning of the XX century
\cite{Cosserat1909},
based on a suggestion by Pierre Maurice Marie \Duhem\ \cite{Duhem1893}.

The proposal was neglected for about fifty years
until brought to attention of the scientific community 
by 
\Truesdell\ and \Toupin\ in \cite{TruesdellToupin1960}
and later reformulated by Ahmed Cemal \Eringen\ in a number of papers
with various possible proposals of micropolar, microstretch and micromorphic
models \cite{Eringen1968a,Eringen1968b}.

As evidenced in \cite{RomanoCMT2016},
all these micro-structured models, whose underlying base continuum
is a standard \Cauchy\ model with geometric dimension greater than one,
are affected by a redundancy of the
implicit description of micro and macro stretching.

Redundancy means the set of scalar conditions
expressing vanishing of the rate of deformation
is not minimal.
Redundancy is due to requirements of kinematic compatibility,
as shown for 3D models by relying on mathematical
results of integrability and regularity of solution 
\cite{RomanoCMT2016,Treatise2023}.

Elimination of redundancy is a challenging task 
not even attempted 
by scholars engaged with the various polar models conceived by \Eringen\
since nobody had the idea of checking this well-posedness condition,
although it is well-known to researchers in constrained optimisation theory.

But redundancy is responsible for the abnormal proliferation of stress-like parameters
acting as \Lagrange\ multipliers of implicit constraint expressing the
vanishing of the rate of deformation 
adopted by the model.
In this respect, the classical \Euler-\Cauchy\ model of fluid and solid continua
stands a champion of optimality due physical soundness
and simplicity consequent to non-redundancy.

An exception is the simplest non-redundant micromorphic model proposed by the
authors in \cite{RomanoCMT2016}.

\section{Referential formulations in Dynamics: a diabolic deceit}
\label{sec: refDyn}

Alleged treatments of dynamical equilibrium in terms of fields 
on a reference avatar manifold,
are developed in literature under the term \Lagrange\ formulation.

However, the proposed models for 3D continua are
in blatant contrast with the correct notion of equilibrium 
expressed in terms of interaction between force and rigid virtual velocities 
acting and based on the current configuration of the body.

This original idea about equilibrium was first enunciated 
by Johann \Bernoulli\ in his famous letter to Pierre \Varignon,
and formulated by 
\Dalembert\ \cite{dAlembert1743}
and \Euler\ \cite{Euler1744} 
about thirty years later and subsequently extended to continua,
as here shown in modern terms by Eqs.\eqref{fm: Dalembert} and \eqref{fm: Eulerdyn}.

According to \Truesdell\ and \Toupin\ in
\cite[\msec{210}, p.553]{TruesdellToupin1960},
in the notes to the section entitled 
\emph{The equations of motion expressed in terms of a reference state},
Gabrio \Piola\ \cite{Piola1833,Piola1836}
was the first to conceive and propose referential formulations of equilibrium.

The same diabolic suggestion was later experienced 
by Gustav \Kirchhoff\ \cite{Kirchhoff1852,Kirchhoff1876},
Carl Gottfried \Neumann\ \cite{NeumannCG1860}
and Eug{\`e}ne \&\ Fran{\c c}ois \Cosserat\ \cite{Cosserat1909}.

The topic was also debated by the Italian school of Rational Mechanics
leaded by Antonio \Signorini\ \cite{Signorini1930},
Carlo \persone{Tolotti} \cite{Tolotti1943},
Francesco \persone{Stoppelli} \cite{Stoppelli1954}
and Giuseppe \persone{Grioli} \cite{Grioli1961}
in the period about World War II.

In the encyclopedic articles  
\cite[\msec{210}, p.553]{TruesdellToupin1960}
and \cite[\msec44, p.127]{TruesdellNoll1965},
drawn up under guidance of  \Truesdell,
these contributions are quoted and commented upon.

Since then, many (we could even dare to say all) scholars in \CM\ 
have reiterated the \emph{mechanical crime}
of referential formulations of equilibrium.

In \cite[Eq.(44.12--15)]{TruesdellNoll1965},
it is also quoted that, according to Antonio \Signorini\ \cite{Signorini1930,Signorini1933},
the referential condition of rotational equilibrium 
of the reshaped boundary traction and bulk actions
has to be considered as a \emph{compatibility condition}
to be verified \emph{a posteriori},
that is once the placement map is made available by 
evaluation of the \emph{deformed} configuration.

The proposed reshaping, consisted of a 
rescaling according to ratios of surface areas and bulk volumes
and by distant parallel transport by translation
to the new base points 
on the reference configuration
as described in \cite{TruesdellNoll1965},
revisited in \cite{Gurtin1981,Podio2000}
and recently addressed in \cite{Spacetime2023}.

A simple comparison between the pertinent expressions
reveals the condition of translational equilibrium
in terms of reshaped reference forces 
is by construction
equivalent to the original one.

According to
\cite[\msec44, p.127]{TruesdellNoll1965}
the condition of rotational equilibrium 
is expressed by a hybrid formulation in which
the radii from the pole are attached to the original force base points 
in the current configuration but expressed as function of the
corresponding reference points by means of the placement map.

Coincidence with the original rotational equilibrium
condition is thus achieved but any usefulness of the
procedure is lost.
Contributions in literature are completely silent about the 
crucial question of how to perform the reshaping
of constraints to be imposed on a reference potato-shaped avatar.
This deficiencies deprive the entire procedure of 
mechanical significance and engineering interest.

Following 
\cite[\msec{210}, p.553]{TruesdellToupin1960},
\Gurtin's fathomless opinion in \cite[Ch.7]{Gurtin1983} was:\\
\emph{In many problems of interest
—especially those involving solids—
it is not convenient to work with $\,\BT\,$,
since the deformed configuration is not known in advance.}

In this assertion $\,\BT\,$
denotes the \Cauchy\ \emph{true} stress tensor:
\begin{equation}
\BT\in\MIX\di{\Tconf}:\TT\conf\mapsto\TT\conf\,.
\label{fm: }
\end{equation}
The proposal was to resort to the (first) \Piola\ tensor:
\begin{equation}
\piola\equaldef\BT\punto\cof\di{\BF_\pp}:\TT\confref\mapsto\TT\conf\,.
\label{fm: Piola}
\end{equation}
The relation in Eq.\eqref{fm: Piola}
depends on the placement map
$\,\pp:\confref\mapsto\conf\,$ 
from the reference configuration to the unknown configuration $\,\conf\,$
where equilibrium has to be checked,
through the tangent map:
\begin{equation}
\BF_\pp\equaldef\TT\pp:\TT\confref\mapsto\TT\conf\,,
\label{fm: }
\end{equation}
the cofactor operator being defined by:
\begin{equation}
\cof\di{\BF_\pp}\equaldef
\det\di{\BF_\pp}\punto\BF_\pp^{-A}:\TT\confref\mapsto\TT\conf\,.
\label{fm: }
\end{equation}
Further details are given in Appendix \ref{sec: piola}.

Resorting to \Piola\ tensors entails 
\emph{falling out of the pan directly into the fire}.

In fact, the configuration $\,\conf\,$, where equilibrium has to be imposed,
is not the known starting configuration 
(assumed in equilibrium) of the incremental process.

Rather $\,\conf\,$ is the next-to-be-detected equilibrium configuration
acted upon by the incremental data updated by the loading and shape controlling algorithm.

This algorithm in fact takes into account the trial movement, the imposed shape 
and loading modifications
and performs the ensuing correction required by the equilibrium gap control.

A fine example of the matter is provided by the dynamical analysis 
of a sailing boat during a challenge round. 
Here the main loading is exerted by the action of
the wind on the sail and by the water on the boat hull, rudder and keel
and depends on the varying position 
of the sail and of the rudder with respect to the boat and on the varying relative
direction between boat and wind.
Other striking examples are provided by a bike ride and by 
the simpler case of an acrobat walking on a wire.

In all applicative analyses the incremental displacement stands therefore 
a priori unknown.

Knowledge of a reference placement 
is illusory and moreover also
the correspondence between 
the final and the reference configuration is completely unknown,
with an infinite number of candidates ready to get the role of positioning map
and no selection and construction criteria available.

Well-posedness requires the result of this procedure to be
independent of the choice of the placement map.
In \cite[(13.1)]{Podio2000} the reference shape is declared 
to have been chosen \emph{once and for all}.
But rationale and method to perform this choice are not discussed, 
nor the issues ensuing from it.

The target of any structural analysis 
is the evaluation of an exact (or approximate) displacement field 
yielding the solution of the relevant incremental equilibrium problem
consequent to a step forward of the data control algorithm.

Therefore, the fact that the \emph{deformed configuration} 
is \emph{a priori unknown}
is an unavoidable feature of equilibrium problems
in the realm of large displacements.

The request of having complete information available
on the final result from the very beginning of the nonlinear procedure
requires a supernatural foresight,
with the consequence of rendering inconsistent all structural methods of nonlinear analysis 
which are actually based on iterative trial and error algorithms.

\section{Non-Linear Dynamics}
\label{sec: goal}

The ultimate goal of nonlinear structural analysis
is the evaluation of the \emph{unknown} movement 
$\,\moto_\lapse:\conf_0\mapsto\TjE\,$
subsequent to a prescribed process of actions exerted on the structural model
(additional forces, impressed movements, variations of 
electric, magnetic and temperature field, etc.), 
starting from the current known configuration $\,\conf_0\,$ in dynamical equilibrium.

The relevant trial and error procedure for an elastostatics problem 
is outlined in \msec \ref{sec: EVOL} below.

The diabolic deception underlying proposals of
referential formulations of equilibrium
was unveiled in
\cite{Treatise2023} and further investigated in 
\cite{Spacetime2023,Advancements2024}.

Prior to quotation by \Truesdell\ and \Toupin\ 
in \cite[\msec{210}, p.553]{TruesdellToupin1960},
the count Gabrio \Piola\ \persone{Daverio}
was essentially a \persone{Carneades} %
\footnote{{\ }%
\persone{Carneades} of \persone{Cyrene} (214--129 B.C.)
ancient greek philosopher,
head of the Skeptical Academy in Athens.
The way of saying 
"\persone{Carneades}! Who was this guy?"
is borrowed from the novel 
\emph{I promessi sposi (The Betrothed)} (1827) by 
Alessandro \persone{Manzoni} (1785--1873)
and since then stands for a \emph{completely unknown person}.
}
in the \CM\ community of the XX century.
The responsibility of a non-critical revisitation of his formulae
and of dissemination in the \CM\
has to be taken mainly by the authors of 
\cite{TruesdellToupin1960,TruesdellNoll1965}
and their followers.

In light of these considerations, the recent flowering of 
monographs 
dedicated by Italian scholars
to Gabrio \Piola\ contributions to \CM,
\cite{CapecchiRuta2007,CapecchiRuta2011,Capecchi2012} and
\cite{PiolaCW2014,PiolaCW2018},
appears largely inadequate,
due to serious errors so disseminated.

With any evidence, editors and authors of these volumes
fell themselves victims of the same deceptive devil
who sneakily suggested feasibility of formulations of equilibrium in terms of
a referential placement.

On the other side, skilled structural engineers should be 
equipped with cultural weapons which, 
on the basis of the original investigations and of the grand ideas 
conceived by the Fathers of \CM,
are effective in defeating these temptations.

What seems to be attributed to \Piola\ \cite{Piola1833}
is the merit of having suggested to make recourse to
the method of \Lagrange\ multipliers as suitable tool 
in defining the stress field in duality with the rigidity constraint
on spatial virtual velocities.
This merit was remarked in 
\cite{TruesdellToupin1960}
and substantiated here by Eq.\eqref{fm: finn}.

As explicated in \cite{RomanoDiaco2004},
the modern mathematical result,
underlying the method of \Lagrange\ multipliers for introducing the stress field in 
3D \CM,
is the fundamental  \emph{closed range theorem} of Functional Analysis,
due to Stefan \Banach\ \cite{Banach1932}.

On the contrary, introduction of first and second
\Piola-\Kirchhoff\ tensors,
so named by \Truesdell\ and \Toupin\ 
in \cite[\msec{210}, p.553]{TruesdellToupin1960},
should not be quoted as merits but rather forgiven, forsaken and forgotten
as notions involved in persisting misconceptions concerning equilibrium
and constitutive relations
\cite{Innsbruck2014,Spacetime2023,Advancements2024}.

\section{Computational Dynamics}
\label{sec: CompDyn}

Let us now discuss,
with reference to the context of automatic structural computations,
the applicative relevance 
of the critical observations brought about above.

\subsection{Alleged Lagrange vs Euler formulations}
\label{sec: LagDyn}

The so called \Lagrange\ (or referential) formulation 
of structural problems,
(to be compared with the so called \Euler\ spatial formulation)
has been proposed also for computational tasks.

This misleading nomenclature, introduced in the context of Fluid-Dynamics,
is not supported, neither by historical evidence nor by
mathematical reasoning or applicative usefulness, and
should profitably be amended.

The issue was dealt with in \msec\ref{sec: windows}
in discussing about traveling control windows.

What is named after \Lagrange\ is the basilar law of motion
imposed on the current configuration
of a massive body with convectively conserved mass-form along the motion
as illustrated by Eq.\eqref{fm: twocond}.

On the other hand, what is named after \Euler\ is an application of
the additivity property of differential calculus to get the split in
Eq.\eqref{fm: accvelsplit}
when both the involved derivatives are feasible.
Unfeasibility may occur at the boundary of 3D body  
and anywhere in lower dimensional bodies,
such as bullets, wires and membranes.

\subsection{Updated Lagrange formulations}
\label{sec: UpLagDyn}

In Computational Dynamics,
updated \Lagrange\ formulations have been also proposed
especially when, in large displacement 
Finite Element Method (\FEM) computations, 
severely distorted meshes are needed to be repaired.

The updating consists of taking  
the configuration of the body, at the beginning of
each time-step in the incremental process, as reference manifold.

Both of these proposals,
the formulation named after \Lagrange\ and its updating,
are replicated even
in recent treatments of computational mechanics by
Mike \Crisfield\
\cite{Crisfield1996},
Ted \Belytschko, Wang Kam \Liu\ \&\ Brian \Moran\
\cite{Belytschko2001}
and by
René \Borst\ et al.
\cite{deBorst2012}.

To these treatments the critical comments expressed 
in the last two sections are still applicable
since, contrary to improper referential formulations,
the correct computational procedure consists in imposing
the equilibrium condition on the available current estimate of the 
body configuration under the updated data provided by the control algorithm,
as will be detailed in the next \msec\ref{sec: EVOL}.

\subsection{Evolutive equilibrium}
\label{sec: EVOL}

This section, renewed from \cite{Spacetime2023},
provides a brief sketch of a stepwise iterative procedure
for the solution of an incremental elastostatic problem.

Let us consider at time $\,\ttt\in\ZEIT\,$ 
a body in the spatial configuration $\,\conf\,$
whose kinematics is defined by a linear space 
$\,\cV_\conf\,$
of piecewise regular spatial vector fields 
$\,\Bv_\conf:\conf\mapsto\TT\EUconf\,$,
being $\,\EUconf\,$ the spatial slice containing $\,\conf\,$.

\goodbreak

The body is assumed to be subject to affine kinematic constraints
described by a linear subspace $\,\cL_\conf\subset\cV_\conf\,$  
of \emph{conforming} kinematic fields
and by an imposed kinematic field $\,\Bw_\conf\in\cV_\conf\,$,
under the action of a force system
$\,\Bf_\conf\in\cF_\conf=\dual{(\cV_\conf)}\,$ 
and of the reactive system 
$\,\Br_\conf\in\cL_\conf^\circ\subseteq\dual{(\cV_\conf)},$
exerted by the affine  constraints.

Accordingly, admissible kinematic fields at $\,\conf\,$
belong to 
the affine variety $\,\cA_\conf\equaldef\Bw_\conf+\cL_\conf\,$. 

In a time lapse $\,\lapse\in\ZEIT\,$
the movement
$\,\moto_\lapse:\TjE\mapsto\TjE\,$
along the trajectory $\,\TjE\subset\EVE\,$
drawn by the body motion in spacetime,
is governed by a control system which
prescribes increments 
of the force system $\,\BDelta\Bf_\conf\in\dual{(\cV_\conf)}\,$
and of driven displacement field $\,\BDelta\Bw_\conf\in\cV_\conf\,$
and also yields the update of the conforming kinematic subspace
from the initial $\,\cL_\conf\subset\cV_\conf\,$
to the final one 
$\,\cL_{\moto_\lapse\di\conf}\subset\cV_{\moto_\lapse\di\conf}\,$,
after the time lapse $\,\lapse\,$.

Let us assume for simplicity a smooth quasi-static evolution 
with the material not leaving the elastic range.

A initial guess 
on the realisation of the body configuration and of the control upgrade
at the end of the incremental step can be performed by evaluating
the increment $\,\BDelta\Bu\in\BDelta\Bw_\conf+\cL_\conf\,$
of displacement from the configuration $\,\conf\,$,
associated with the data increment $\,\set{\BDelta\Bf_\conf,\BDelta\Bw_\conf}\,$ 
at the beginning of the time-step.

This evaluation of the incremental displacement
is based on the rate formulation of equilibrium
\cite{Rate2014,Genesis2022}
in which test fields are assumed to be parallel transported by the motion
along the trajectory, in accord with \Euler\ law of Dynamics.

Replacing time rates with finite increments,
the Rate Virtual Power Principle (\RVPP) takes the form
of an Incremental Virtual Power Principle (\IVPP):%
\begin{equation}
\setlength{\jot}{8pt}
\begin{aligned}
\scalar{\BDelta\Bf}{\dvect}_\conf
&\,=\integrale{\conf}{}\scalar{\BDelta\BSigma}{\deform\di\dvect}\,\massform
+\integrale{\conf}{}\scalar{\BSigma}{\BDelta\deform\di{\BDelta\Bu,\dvect}}\,\massform\,.
\end{aligned}
\label{fm: RVPP}
\end{equation}

This variational principle holds
for any $\,\dvect\in\cL_\conf\,$,
being:
\begin{equation}
\setlength{\jot}{8pt}
\left\{
\begin{aligned}
&\,\BDelta\massform\equaldef\Lieder_{(\BDelta\Bu)}\massform=\Bzero\,, \quad
\textrm{by conservation of mass}\,,
\\
&\,\BDelta(\dvect)\equaldef\nabla_{(\BDelta\Bu)}\dvect=\Bzero\,,\quad\textrm{by construction}\,.
\end{aligned}
\right.
\label{fm: name}
\end{equation}
The mixed incremental stretch tensor 
$\,\BDelta\deform\di{\vel,\dvect}\,$, associated with
the covariant tensor:
\begin{equation}
\unmezzotext\,\Lieder_{\BDelta\Bu}\Lieder_{\dvect}\di\metric
=\metric\punto\BDelta\deform\di{\BDelta\Bu,\dvect}\,,
\label{fm: deformrate}
\end{equation}
is evaluated to be \cite{Rate2014}:
\begin{equation}
\BDelta\deform\di{\BDelta\Bu,\dvect}=\sym_\metric\Bigdi{(\nabla\BDelta\Bu)^A\punto\nabla\dvect}\,.
\label{fm: ratedef}
\end{equation}
Elasticity is expressed by the incremental constitutive relation:
\begin{equation}
\BDelta\BE=\BH\di\BSigma\punto\BDelta\BSigma\,.
\label{fm: }
\end{equation}

The tangent elastic compliance $\,\BH\di\BSigma\,$
is evaluated by the Ludwig Otto \Hesse\ operator of a 
smooth strictly convex scalar stress potential $\,\BXi\,$:%
\footnote{{\ }%
The fiber derivative $\,\derF\,$ is taken along spatial directions
at the pertinent time instant.}%
\begin{equation}
\BH=\derF^2\BXi\,.
\label{fm: }
\end{equation}

By strict convexity $\,\BH\di\BSigma\,$
is positive definite, hence invertible, and such is
the elastic stiffness $\,\BK\di\BSigma=\inv{\BH\di\BSigma}\,$
so that:%
\begin{equation}
\BDelta\BSigma=\BK\di\BSigma\punto\BDelta\BE\,.
\label{fm: elastiff}
\end{equation}

In a purely elastic range:
\begin{equation}
\BDelta\BE=\deform\di{\BDelta\Bu}\,.
\label{fm: pelast}
\end{equation}

The incremental elastic equilibrium problem consists in searching for
an admissible displacement
$\,\BDelta\Bu\in\BDelta\Bw_\conf+\cL_\conf\subset\cV_\conf\,$
fulfilling, for all conforming test fields $\,\dvect\in\cL_\conf\,$, the variational condition:
\begin{equation}
\setlength{\jot}{8pt}
\begin{aligned}
\scalar{\BDelta\Bf}{\dvect}_\conf
&\,=\integrale{\conf}{}
\scalar{\BK\di\BSigma\punto\deform\di{\BDelta\Bu}}{\deform\di\dvect}\,\massform
\\
&\,+\integrale{\conf}{}\scalar{\BSigma}{\BDelta\deform\di{\BDelta\Bu,\dvect}}\,\massform\,.
\end{aligned}
\label{fm: ELARVPP}
\end{equation}

Provided the static variational problem Eq.\eqref{fm: ELARVPP}
admits a displacement solution $\,\BDelta\Bu\,$,
a first trial $\,\moto_\lapse\di\conf\,$ for the deformed configuration is available.%
\footnote{{\ }%
In general a dynamical analysis is needed to ensure existence.
Uniqueness breaks down when the incremental elastic response becomes singular
and instability phenomena take place.
Early computational investigations about stability and accuracy 
were provided in \cite{Newmark1959,Casciaro1975}.
}

Then, with $\,\BDelta\BSigma\,$ 
given by Eq.\eqref{fm: elastiff}-Eq.\eqref{fm: pelast},
 the incremental elastic response
$\,\BDelta\Br_{\moto_\lapse\di\conf}\in\cF_{\moto_\lapse\di\conf}\,$ is given by:%
\begin{equation}
\setlength{\jot}{8pt}
\begin{aligned}
&\,\scalar{\BDelta\Br}{\dvect}_{\moto_\lapse\di\conf}
\\
&\,=\integrale{\moto_\lapse\di\conf}{}
\moto_\lapse\push\scalar{\BK\di{\BSigma+\BDelta\BSigma}\punto\deform\di{\BDelta\Bu}}{\deform\di\dvect}\punto\massform
\\
&\,+\integrale{\moto_\lapse\di\conf}{}
\moto_\lapse\push\scalar{\BSigma+\BDelta\BSigma}{\BDelta\deform\di{\BDelta\Bu,\dvect}}\punto\massform\,,
\end{aligned}
\label{fm: RVPPtest}
\end{equation}
since $\,\moto_\lapse\push\massform=\massform\,$
by conservation of mass along the motion:

The response in Eq.\eqref{fm: RVPPtest} is compared with the incremental force:
\begin{equation}
\scalar{\BDelta\Bf}{\dvect}_{\moto_\lapse\di\conf}\,,
\label{fm: }
\end{equation}
in which $\,\dvect_{\moto_\lapse\di\conf}\,$ is parallel transported
by the movement $\,\moto_\lapse:\TjE\mapsto\TjE\,$:
\begin{equation}
\dvect_{\moto_\lapse\di\conf}\equaldef\moto_\lapse\forw\,\dvect_{\conf}\,.
\label{fm: partras}
\end{equation}
Here $\,\forw\,$ denotes the forward distant parallel transport 
in the \Euclid\ space $\,\EU\,$
from the spatial configuration $\,\conf\,$
to the displaced one $\,\moto_\lapse\di\conf\,$.
At this stage the force increment 
$\,\BDelta\Bf_{\moto_\lapse\di\conf}\in\cF_{\moto_\lapse\di\conf}\,$ 
is updated by the control system.

The incremental equilibrium gap 
$\,\BDelta\Br_{\moto_\lapse\di\conf}-\BDelta\Bf_{\moto_\lapse\di\conf}\,$
is applied to the trial configuration 
$\,\moto_\lapse\di\conf\,$ corresponding to the running iteration.

The elastic equilibrium
Eq.\eqref{fm: ELARVPP}
with $\,\moto_\lapse\di\conf\,$ taking the place of $\,\conf\,$,
updates the current guess and another iteration for 
the elastic incremental displacement solution is performed.
The iterative loop comes to a stop provided that a suitably chosen norm of
the equilibrium gap becomes smaller than a prescribed tolerance.

\section{Concluding remarks}
\label{sec: conclusion}

After so many years of persistence of improper formulations of Frame-Changes,
Material Frame Indifference, Equilibrium in terms of a reference placement
and modelling of constitutive behaviour according to a 
chain (multiplicative) scheme of Elasto-Anelasticity,
all these misstatements could 
with good reason be deemed devil suggested horrors.
%
\footnote{
This terminology was suggested to the senior author by 
a remembrance of the years 1956--1958 spent 
by the first author with his twin brother Manfredi
at Classical Lyceum \emph{Umberto I} in Naples (Italy), 
and especially of a clever and demanding teacher of Chemistry 
prof. Anna Rippa,
whose frequent exclamation was: "But this is not an error, it is an horror!".
The claim was funny due to big difficulties of the theacher  
in pronouncing the rolled consonant
"r" in Italian.
}

The formulation developed in \cite{TruesdellToupin1960} and \cite{TruesdellNoll1965} by well-respected scholars, having been replicated and exemplified by many followers
\cite{Gurtin1981,OdenReddy1982,MarsdenHughes1983,Ogden1984,Podio2000},
are nowadays spread in the global community of
Continuum Mechanics.

Incorrectness of the procedure of referential equilibrium 
and of any occurrence of \emph{reference shapes} in Continuum Mechanics
requires an emergency act to avoid damages 
to structural mechanics and engineering design, 
especially if put into operation within automatic computational codes.

The proposed rate model of elastic and anelastic constitutive response 
of involved materials provides an effective tool of analysis
able to eliminate \emph{ab initio} the remarked issues.
Application of the new elasticity rate theory to trusses can be found in \cite{BarrettaVU_trusses_MRC2025}.

\bmhead{Acknowledgements}
Financial support from the Italian Ministry of University and Research (MUR) 
in the framework of the Project PRIN 2022 Nonlocal Mechanics of Innovative Soft Nanostructures (code 2022ZW2NMJ) funded by the European Union - 
Next Generation EU is gratefully acknowledged. 

\begin{figure}[h]
\centering
\setlength{\fboxsep}{1.5mm}
\qquad
\begin{subfigure}{0.40\textwidth}							
\fbox{\centering			
\includegraphics[width=0.40\textwidth]{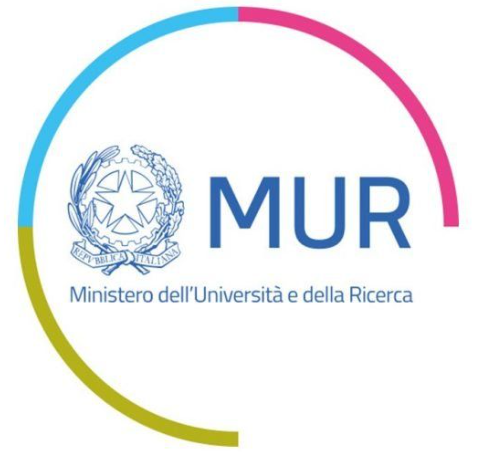}}
\end{subfigure}		
\qquad
\begin{subfigure}{0.40\textwidth}							
\fbox{\centering			
\includegraphics[width=0.80\textwidth]{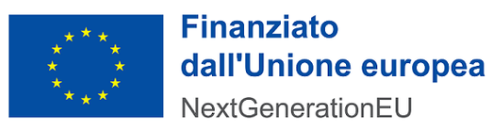}}		
\end{subfigure}		
\end{figure}

\noindent\textbf{Author contributions}
The authors contributed to the work in equal measure.
\section*{Declarations}

\noindent\textbf{Competing interests}
The authors declare no competing interests.




\section{Appendix}
\label{sec: piola}

Let us refer to definitions given in \msec\ref{sec: refDyn}.
Alleged formulations of equilibrium in terms of a reference placement
start from
\Euler-\Jacobi\ formula for volumetric expansion:%
\begin{equation}
\map\pull\volformg=\JJ_{\refmap}\punto\volformg\,,
\label{fm: EulerJacob}
\end{equation}
with $\,\volformg\,$ metric volume form in \Euclid\ space with metric $\,\metric\,$
and $\,\JJ_{\refmap}\equaldef\det\di{\BF}\,$.

The spatial unit normals
$\,\Bn_{\pconfref}:\pconfref\mapsto\TT_{\confref}\EU\,$,
to the reference boundary surface $\,\pconfref\,$
and
$\,\Bn_{\pconf}:\pconf\mapsto\TT_{\conf}\EU\,$
to the current boundary surface $\,\pconf\,$
enter in the definitions of the area-forms:
\begin{equation}
\setlength{\jot}{8pt}
\begin{aligned}
\volform_{\pconf}&\,\equaldef\volformg\punto\Bn_{\pconf}\,,\quad\textrm{on }\pconf\,,
\\
\volform_{\pconfref}&\,\equaldef\volformg\punto\Bn_{\pconfref}\,,\quad\textrm{on }\pconfref\,.
\end{aligned}
\label{fm: areas}
\end{equation}

The volume of the parallelepiped $\,\cP\,$
generated by the unit normal $\,\Bn_{\pconf}\,$
over the base area $\,\volform_{\pconf}\,$ on $\,\pconf\,$,
is given by tensor product decomposition:
\begin{equation}
\setlength{\jot}{8pt}
\begin{aligned}
\volform_{\cP}
&\,=\di{\metric\Bn_{\pconf}}\otimes\volform_{\pconf}\,.
\end{aligned}
\label{fm: decomp}
\end{equation}
In the third order tensor $\,\volform_{\cP}\,$,
the first argument is transversal
and the last two are tangent to $\,\pconf\,$:
\emph{the volume of the parallelepiped 
is height times base-area}.

Application of \Euler-\Jacobi\ transformation Eq.\eqref{fm: EulerJacob} 
yields directly Edward John \Nanson\ formula \cite{Nanson1878}:
\begin{equation}
\refmap\pull\Bigdi{\di{\metric\Bn_{\pconf}}\otimes\volform_{\pconf}}
=\JJ_{\refmap}\punto\Bigdi{\di{\metric\Bn_{\pconfref}}\otimes\volform_{\pconfref}}\,.
\label{fm: EULJ}
\end{equation}

The expression in terms of the cofactor 
\begin{equation}
\cof\di{\BF}\equaldef
\JJ_{\refmap}\punto\BF^{-A}:\TT\confref\mapsto\TT\conf\,,
\label{fm: cofactor}
\end{equation}
is got by the equality:
\begin{equation}
\setlength{\jot}{8pt}
\refmap\push\bigdi{\JJ_{\refmap}\punto\metric\Bn_{\pconfref}}
=\metric\punto\di{\cof\di{\BF}\punto\Bn_{\pconfref}}\,.
\label{fm: forinter}
\end{equation}
To prove Eq.\eqref{fm: forinter} it suffices to observe that
for any $\,\vect\in\VV_{\conf}\EVE\,$:%

\newcommand{\Bnref}{\Bn_{\pconfref}}
\newcommand{\JJref}{\JJ_{\refmap}}

\begin{equation}
\setlength{\jot}{8pt}
\left\{
\begin{aligned}
\scalar{ \refmap\push\di{\metric\Bnref} } {\vect}
&\,=\refmap\push\scalar{ \metric\Bnref }{\refmap\pull\vect}
\\
&\,=\refmap\push\scalar{ \metric\Bnref }{\inv{\BF}\vect}
\\
&\,=\scalar{ \metric\punto\BF^{-A}\Bnref }{\vect}\,.
\end{aligned}
\right.
\label{fm: name}
\end{equation}
Hence, by arbitrariness of $\,\vect\in\VV_{\conf}\EVE\,$:
\begin{equation}
\refmap\push\di{\metric\Bnref}
=\metric\di{\BF^{-A}\Bnref}\,,
\quad\textrm{on}\;\pconf\,,
\label{fm: pushgn}
\end{equation}
so that the push of Eq.\eqref{fm: EULJ} by $\,\placement:\confref\mapsto\conf\,$ 
yields \Nanson\ formula:
\begin{equation}
\di{\metric\Bn_{\pconf}}\otimes\volform_{\pconf}
=\metric\punto\di{\cof\di{\BF}\punto\Bnref}\otimes\di{\refmap\push\volform_{\pconfref}}\,.
\label{fm: EULJpush}
\end{equation}

Despite the esoteric appearance of the cofactor map,
Eq.\eqref{fm: EULJpush} is just a rewriting,
for the special parallelepiped in Eq.\eqref{fm: decomp},
of \Euler-\Jacobi\ formula
Eq.\eqref{fm: EulerJacob}
for the volumetric expansion.

The very introduction of the cofactor map as gradient of the determinant 
function is also due to \Euler\ \cite{Treatise2023}:
\begin{equation}
\nabla\di{\det\BF}=\cof\di\BF\,.
\label{fm: Eulcof}
\end{equation}

Given any pair of non parallel tangent vectors $\,\Ba,\Bb\in\TT\pconfref\,$,
tangent to $\,\pconfref\,$ the generated parallelogram
and the pushed one with sides
$\,\BF\Ba,\BF\Bb\in\TT\pconf\,$
have areas $\,{\area}_{\confref}\,$ and $\,{\area}_{\conf}\,$,
given by:
\begin{equation}
\setlength{\jot}{8pt}
\begin{aligned}
{\area}_{\confref}&\,\equaldef\volform_{\pconfref}\coppia{\Ba}{\Bb}\,,
\\
{\area}_{\conf}&\,\equaldef\volform_{\pconf}\coppia{\BF\Ba}{\BF\Bb}\,.
\end{aligned}
\label{fm: areapush}
\end{equation}

The original \Nanson\ formula,
as reported in \cite[Eq.(20.8)]{TruesdellToupin1960}
and \cite[Eq.(22.2.18)]{Ogden1984},
may be expressed in geometric terms as follows
\cite{Spacetime2023}:%
\begin{equation}
{\area}_{\conf}\punto\Bn_{\pconf}
=\cof\di{\BF}\punto\di{ {\area}_{\pconfref}\punto\Bn_{\pconfref}}\,,
\quad\textrm{on}\;\pconf\,.
\label{fm: Nansontre}
\end{equation}

In terms of \Cauchy\ stress $\,\BT:\Tconf\mapsto\Tconf\,$,
the boundary traction is given by the well known formula:
\begin{equation}
\Bt\equaldef\BT\punto\Bn_{\pconf}:\pconf\mapsto\EUconf\,.
\label{fm: }
\end{equation}

Then \Nanson\ formula Eq.\eqref{fm: Nansontre}  
leads to the evaluation:
\begin{equation}
\setlength{\jot}{4pt}
\begin{aligned}
\Bt\punto\area_{\pconf}
&\,=\di{ \BT\punto\Bn_{\pconf} }\punto\area_{\pconf}
\\
&\,=\BT\punto\cof\di{\BF}\punto\di{ \Bn_{\pconfref}\punto\area_{\pconfref} }
\\
&\,=\di{\piola\punto\Bn_{\pconfref}}\punto\area_{\pconfref}\,.
\end{aligned}
\label{fm: piolahomo}
\end{equation}

This correspondence deceptively suggested
the possibility of viable paths towards a formulation
of equilibrium conditions in terms of a reference placement $\,\confref\,$
\cite{Piola1833,Piola1836}.
At this point, two different interpretations are offered in literature.

The former interpretation, according to 
\cite[\msec{210}, p.553]{TruesdellToupin1960}
and \cite[\msec44, p.127]{TruesdellNoll1965}
is no more than a repetition of cardinal equations of Statics
evaluated in the current configuration but written in terms of referential coordinates.

The latter interpretation,
see e.g. \cite[\msec{1.2.4}]{NguyenQS2000},
pretends to impose the equations of equilibrium in the reference configuration
by parallel translation of surface tractions and body forces.

The former approach results in a useless complication.
The latter is an impossible task because 
the system of parallel translated forces 
doesn't necessarily fulfil the cardinal equations of rotational equilibrium
in the reference configuration.
A further and fatal difficulty is met in trying to write the kinematic constraint 
conditions in terms of referential fields.

\end{document}